\documentclass[twocolumn,trackchanges]{aastex701}
\usepackage{lmodern}
\usepackage{comment}
\usepackage{amsmath}

\usepackage{placeins}

\usepackage{float}
\usepackage{placeins}

\renewcommand{\thefootnote}{\fnsymbol{footnote}}

\usepackage{booktabs}
\usepackage{tabularx}
\usepackage{array}

\newcolumntype{C}{>{\centering\arraybackslash}X}

\usepackage{fontspec}
\usepackage[english]{babel} 
\babelprovide{punjabi}
\babelfont[punjabi]{rm}[Renderer=Harfbuzz,Script=Gurmukhi]{FreeSerif.otf}
\newcommand{\Punj}[1]{\foreignlanguage{punjabi}{#1}}

\usepackage{xeCJK}
\begin{document}

\title{JWST Spectra Conclusively Show an Excess of Neutral Gas Outflows in Quiescent Galaxies at z=2--5}

\author[orcid=0009-0000-7075-5554]{Rion Oh (오리온)}
\affiliation{Department of Astronomy, University of Washington, Physics-Astronomy Building, Box 351580, Seattle, WA 98195-1700, USA}
\affiliation{Department of Physics, KAIST, Daejeon 34141, Republic of Korea}
\email[show]{oro020@kaist.ac.kr}  

\author[0000-0002-3475-7648]{Gourav Khullar (\Punj{ਗੌਰਵ ਖੁੱਲਰ})~$^\S$}
\altaffiliation{Baum Postdoctoral Fellow for Innovative Astronomy}
\affiliation{Department of Astronomy, University of Washington, Physics-Astronomy Building, Box 351580, Seattle, WA 98195-1700, USA}
\affiliation{The DiRAC Institute, University of Washington, Physics-Astronomy Building, Box 351580, Seattle, WA 98195-1700, USA}
\affiliation{eScience Institute, University of Washington, Physics-Astronomy Building, Box 351580, Seattle, WA 98195-1700, USA}
\email{gkhullar@uw.edu}

\author[0000-0002-7530-8857]{Arianna S. Long~$^\S$}
\affiliation{Department of Astronomy, University of Washington, Physics-Astronomy Building, Box 351580, Seattle, WA 98195-1700, USA}
\email{aslong@uw.edu}

\author[0000-0002-5293-3975]{Julissa Sarmiento}
\affiliation{Department of Physics and Astronomy, University of Pittsburgh, Pittsburgh, PA 15260, USA}
\email{jms840@pitt.edu}

\author[0000-0003-3596-8794]{Hollis B. Akins}
\email{hollis.akins@gmail.com}
\altaffiliation{NSF Graduate Research Fellow}
\affiliation{The University of Texas at Austin, 2515 Speedway Blvd Stop C1400, Austin, TX 78712, USA}

\author[0000-0002-0930-6466]{Caitlin M. Casey}
\email{cmcasey@ucsb.edu}
\affiliation{Department of Physics, University of California, Santa Barbara, Santa Barbara, CA 93106, USA}
\affiliation{Cosmic Dawn Center (DAWN), Denmark}

\author[0000-0001-8551-071X]{Yingjie Cheng}
\affiliation{Department of Astronomy, University of Washington, Physics-Astronomy Building, Box 351580, Seattle, WA 98195-1700, USA}
\email{yingjiec@uw.edu}

\author[0000-0002-9382-9832]{Andreas L. Faisst}
\affiliation{IPAC, California Institute of Technology, 1200 E. California Blvd. Pasadena, CA 91125, USA}
\email{afaisst@ipac.caltech.edu}

\author[0000-0002-3560-8599]{Maximilien Franco}
\email{maximilien.franco@cea.fr}
\affiliation{Université Paris-Saclay, Université Paris Cité, CEA, CNRS, AIM, 91191 Gif-sur-Yvette, France}

\author[0000-0002-0236-919X]{Ghassem Gozaliasl}
\email{ghassem.gozaliasl@gmail.com}
\affiliation{Department of Computer Science, Aalto University, P.O. Box 15400, FI-00076 Espoo, Finland}
\affiliation{Department of Physics, University of Helsinki, P.O. Box 64, FI-00014 Helsinki, Finland}

\author[0000-0002-3301-3321]{Michaela Hirschmann}
\affiliation{Institute of Physics, GalSpec, EPFL, Observatoire de Sauverny, Chemin Pegasi 51, 1290 Versoix, Switzerland}
\affiliation{INAF, Astronomical Observatory of Trieste, Via Tiepolo 11, 34131 Trieste, Italy}
\email{michaela.hirschmann@epfl.ch}

\author[0000-0003-3216-7190]{Erini Lambrides} \altaffiliation{NPP Fellow}
\affiliation{NASA-Goddard Space Flight Center, Code 662, Greenbelt, MD, 20771, USA}
\email{erini.lambrides@nasa.gov}

\author[0000-0002-9883-7460]{Jacqueline E. McCleary}
\email{j.mccleary@northeastern.edu}
\affiliation{Department of Physics, Northeastern University, 360 Huntington Ave, Boston, MA, USA}

\author[0009-0008-5008-4309]{Tiara Anderson}
\email{tiaraa2@uw.edu}
\affiliation{Department of Astronomy, University of Washington, Physics-Astronomy Building, Box 351580, Seattle, WA 98195-1700, USA}

\author[0009-0000-2577-1619]{David C. Andrews}
\email{davida04@uw.edu}
\affiliation{Department of Astronomy, University of Washington, Physics-Astronomy Building, Box 351580, Seattle, WA 98195-1700, USA}

\author[0009-0000-5333-9970]{Dylan Berry}
\affiliation{Department of Astronomy, University of Washington, Physics-Astronomy Building, Box 351580, Seattle, WA 98195-1700, USA}
\affiliation{The DiRAC Institute, University of Washington, Physics-Astronomy Building, Box 351580, Seattle, WA 98195-1700, USA}
\email{dberry04@uw.edu}

\author[0009-0009-9700-1811]{Nguyễn Bình}
\email{ngbinh@uw.edu}
\affiliation{Department of Astronomy, University of Washington, Physics-Astronomy Building, Box 351580, Seattle, WA 98195-1700, USA}

\author[0009-0007-0553-9610]{Elaine Gammon}
\affiliation{Department of Astronomy, University of Washington, Physics-Astronomy Building, Box 351580, Seattle, WA 98195-1700, USA}
\email{laineygammon@gmail.com}

\footnote[4]{These authors contributed equally as co-mentors to the lead author.}


\begin{abstract}

Galaxies exhibit a broad range of star formation activity, from actively star-forming to quiescent systems. Yet, the mechanisms responsible for the rapid shutdown and continued suppression of star formation remain poorly understood, particularly for the quiescent galaxies recently uncovered by JWST at $z>3$. Neutral gas outflows provide a direct tracer of gas removal and regulation in quiescent galaxies. We therefore present a Na\,\textsc{i}\,D $\lambda\lambda5891,5897$ absorption--line stacking analysis of 274 galaxies at $z=2$--5 drawn from the DAWN JWST Archive (DJA) and the \textsc{ember} JWST program, using JWST/NIRSpec medium-resolution grating spectroscopy and NIRCam photometry. Na\,\textsc{i}\,D absorption is detected in all quiescent galaxy stacks spanning $z=2$--5, providing the first statistical evidence that neutral gas outflows are a ubiquitous feature of quiescent galaxies in the early Universe. Compared to the non-quiescent galaxy stacks, the quiescent stacks show mass loading factors higher by 2--4 dex, with mass outflow rates $\dot{M}_{\rm out} \sim 9$--$30\,M_\odot\,{\rm yr}^{-1}$ elevated by $\sim0.5$--$1$ dex and outflow velocities $v_{\rm out} \sim 310$--$570\,{\rm km\,s}^{-1}$ higher by a factor of $\sim2$. The extreme mass loading factor values and declining star formation histories strongly indicate that these outflows cannot be driven by current star formation alone. In the statistically reliable quiescent redshift bins at $2 \leq z < 3$ and $3 \leq z < 4$, the elevated $[\mathrm{N\,II}]/\mathrm{H}\alpha$ and $[\mathrm{O\,III}]/\mathrm{H}\beta$ ratios are consistent with a possible AGN contribution, suggesting that non-stellar feedback may play a role in sustaining quiescence in galaxies at these epochs.

\end{abstract}


\section {INTRODUCTION}
\label{sec:intro}

Understanding what drives the rapid and complete suppression of star formation in massive galaxies is one of the most pressing unsolved problems in current extragalactic research \citep{Man_2018,Merlin_2019,bluck2023galaxyquenchinghighredshift,kimmig2023blowingcandlequenchgalaxies,lovell2023lightreionisationepochsimulations,2025MNRAS.536.2324L,Whitaker_2026}. Galaxies broadly divide into two populations: actively star-forming systems and quiescent systems in which star formation has largely ceased \citep{Williams_2009,Whitaker_2011}. This bimodality is observed throughout cosmic history, with a substantial fraction of massive galaxies already quiescent at $z \sim 2-3$ \citep{1996Natur.381..581D,Onodera_2012,Toft_2014,2016A&A...592A..19C,2017Natur.544...71G,Schreiber_2018,Belli_2019,2020ApJ...889...93V}. These systems assembled the bulk of their stellar mass and quenched within the first two billion years of cosmic history, in most cases over timescales of only a few hundred Myr \citep{Suess_2022, park2024widespreadrapidquenchingcosmic,beverage2024heavymetalsurveyevolution}. More strikingly, pre-JWST studies had already reported candidate populations of massive quiescent galaxies at $z \sim 3$--4, although spectroscopic confirmation remained limited. JWST has since extended this frontier to $z > 4$ and strengthened the evidence that massive galaxies quenched earlier than predicted by many galaxy formation models \citep{Schreiber_2018,Forrest_2020,Carnall_2023,long2023efficientnircamselectionquiescent,Valentino_2023,antwidanso2024fenikssurveyspectroscopicconfirmation,barrufet2024quiescentdustyunveilingnature,degraaff2024efficientformationmassivequiescent,Nanayakkara_2024,Wang__2024,weibel2025rubiesrevealsmassivequiescent,Khullar2026b}. Proposed quenching mechanisms include active galactic nuclei (AGN), stellar-driven winds, virial shock heating of infalling gas, mergers, and environmental processes. However, the relative importance of these channels, and the epochs at which they operate, remain poorly constrained observationally \citep{Faisst_2017,Khullar_2022,bluck2023galaxyquenchinghighredshift,2025MNRAS.536.2324L,Whitaker_2026}.

A central open question is whether gaseous outflows are a direct cause of quenching, or merely accompany a quenching process already underway. Furthermore, it remains unclear whether outflows can persist even after star formation has largely ceased, maintaining quiescence for billions of years thereafter. In principle, outflows driven by AGN activity or intense star formation can directly remove or heat the cold gas in the ISM, cutting off the fuel supply for star formation \citep{Murray_2005,Veilleux_2005,Fabian_2012,Veilleux_2020}. Testing this causal link requires tracing the gas phase that dominates the mass budget of the outflow. Neutral gas outflows traced by the Na\,\textsc{i}\,D $\lambda\lambda$5891, 5897 absorption doublet are particularly critical in this regard, since the neutral phase often carries mass outflow rates one to two orders of magnitude larger than the ionized phase \citep{Roberts_Borsani_2020,Baron_2021,Avery_2022,Davies_2024}. Na\,\textsc{i}\,D outflows have been detected across star-forming, post-starburst, and quiescent galaxies at lower redshifts, and more recently at higher redshifts \citep{Rupke_2005a, Baron_2021, sun2026censusnadtracedneutral}. Notably, neutral gas inflows have also been observed alongside outflows in long-quenched systems \citep{Bevacqua_2026}. This suggests that quenching may be accompanied by continued gas cycling well after a galaxy has reached the quiescent state, rather than occurring as a single event, a process often described as `maintenance-mode' quenching \citep{Beckmann2017, Patil2026}.

The epoch at $z>2$ is when most massive galaxies form and quench \citep{Thomas_2005,2014ARA&A..52..415M}, but testing whether such continued gas cycling occurs at these redshifts has only recently become feasible in the JWST era. At $z > 2$, the Na\,\textsc{i}\,D doublet shifts into the near-infrared, where it is too faint to detect in typical quiescent galaxies with ground-based spectroscopy. Pre-JWST space-based facilities likewise lacked the wavelength coverage and sensitivity needed for such observations. Moreover, spectroscopically confirmed samples of quiescent galaxies at $z > 3$ remained small, because existing data could select plausible candidates but only rarely verify their suppressed star formation \citep{long2023efficientnircamselectionquiescent,Antwi_Danso_2023,Gould_2023,Valentino_2023,Carnall_2023}. As a result, studies of high-redshift neutral gas kinematics were confined solely to a handful of individual objects \citep{Valentino_2025}. Thus, despite emerging evidence for outflows in high-redshift quiescent galaxies \citep{park2024widespreadrapidquenchingcosmic, zhu2026againneutraloutflowsz35}, it remains unclear how ubiquitous neutral gas outflows are after star formation has ceased, and whether their properties can be explained by residual star formation alone.

With its near-IR sensitivities, JWST has now made this test possible for the first time. Recently, a handful of JWST-based studies have begun to report individual detections of Na\,\textsc{i}\,D outflows and inflows in high-redshift galaxies \citep{Davies_2024,taylor2026jwstexcelssurveyoutflows, zhu2026againneutraloutflowsz35,lyu2026statisticaldetectionmgiitracedcool}. However, the signal-to-noise ratio of individual high-redshift spectra is generally insufficient to directly detect the weak Na\,\textsc{i}\,D feature, which requires a high continuum signal-to-noise ratio that is typically achieved only for the most continuum-bright sources \citep{Davies_2024,moretti_empirical_2026}. Consequently, these results remain limited to small samples or individual objects and cannot yet establish how common such outflows are across the star-forming-to-quiescent sequence, or how their properties vary with galaxy type. Spectral stacking offers a direct route around this limitation by constructing a population-averaged residual spectrum in which weak absorption features can be measured statistically, enabling a systematic comparison of outflow signatures across galaxy types and redshift bins.

In this work, we use a spectroscopic sample of 274 galaxies at $z=2$--5 drawn from the DAWN JWST Archive (DJA) and \textsc{ember} JWST survey (GO\#7076, PI: H. Akins) \citep{2025jwst.prop.7076A} to present a statistical search for neutral gas outflows traced by Na\,\textsc{i}\,D absorption across star-forming, intermediate, and quiescent galaxy populations. By stacking continuum-subtracted spectra in bins of galaxy type and redshift, we overcome the signal-to-noise limitations of individual spectra and test whether the presence and strength of Na\,\textsc{i}\,D absorption vary systematically with galaxy evolutionary state and redshift. We further compare Na\,\textsc{i}\,D absorption and inferred outflow properties across galaxy populations, and use mass loading factors, recent star-formation histories, and energy-budget arguments to assess whether residual star formation alone can power the detected outflows.

This paper is organized as follows. Section \ref{sec:method} describes the data, sample selection, stellar population modeling, spectral stacking, line fitting, and outflow-property measurements. Section \ref{sec:results} presents the Na\,\textsc{i}\,D detections and derived outflow properties across galaxy populations and redshift. Section \ref{sec:discussion} discusses the possible driving mechanisms of the observed outflows and their implications for quenching. We summarize our conclusions in Section \ref{sec:conclusion}. Throughout, we adopt a standard flat $\Lambda$CDM cosmology corresponding to WMAP9 observations.

\section{DATA \& METHOD} \label{sec:method}

\subsection{Data Selection and Spectral Preparation}
\label{sec:data}

This analysis combines JWST/NIRSpec medium-resolution spectroscopy with multi-wavelength JWST/NIRCam photometry. The spectroscopic data are drawn from the DAWN JWST Archive (DJA)\footnote{\url{https://dawn-cph.github.io/dja/}} \citep{brammer_2025_15472354} and the \textsc{ember} JWST survey (GO\#7076, PI: H. Akins) \citep{2025jwst.prop.7076A}. The spectra from DJA were reduced using \texttt{msaexp} \citep{brammer_2023_8319596, 2024Sci...384..890H, 2025A&A...697A.189D}, and the imaging mosaics were processed with \texttt{grizli} \citep{brammer_2023_8370018, Valentino_2023}. The \textsc{ember} JWST survey program targets little red dots (LRDs) with JWST/NIRSpec multi-object spectroscopy in the COSMOS, UDS, EGS, and GOODS-S fields, prioritizing the most continuum-bright sources ($m_{\rm F444W}<24$) at $z\sim3-7$ \citep{tanaka2025discoverylittlereddot}. The \textsc{ember} sources used here are filler targets observed on the same masks, rather than the primary LRD targets. MSA pointings are observed with both the PRISM/CLEAR disperser and the G395M/F290LP grating. Data reduction details for the \textsc{ember} JWST survey are presented in detail in Tanaka, Akins, et al. 2026 (in prep), and briefly summarized here: NIRSpec data reduction roughly follows the standard pipeline, starting with products from v1.20.2 of the JWST Calibration Pipeline and a CRDS mapping of \texttt{pmap-1481}. A custom iterative procedure is applied to subtract thermal and 1/$f$ noise from the detector, then flat-field, photometric, and wavelength calibrations are applied to the extracted 2D spectra. 

Across these data, we specifically use spectra obtained with NIRSpec medium-resolution gratings (G140M, G235M, and G395M; $R\sim1000$). We exclude spectra with poor data quality from the analysis, using the visual inspection flags provided by each catalog. For DJA sources, we exclude those with a grade of -1 (Fit not performed or graded), 0 (Spectrum suffers some data quality issue), and \textsc{ember} sources labeled as impossible. We further exclude all spectra whose median per-pixel spectral $S/N$, measured from the 1D spectrum, is below 3. Our spectroscopic data span $2 \leq z < 5$.

For the photometric component of the analysis, we require multi-wavelength photometry for all selected objects in order to perform spectrophotometric SED fitting. For the photometric data, we use JWST/NIRCam data from the COSMOS-Web Survey \citep{casey2023cosmosweboverviewjwstcosmic,Shuntov2025,franco2025cosmoswebcomprehensivedatareduction}, as well as additional archival JWST imaging data from other ancillary surveys, including PRIMER-COSMOS \citep{2021jwst.prop.1837D,Donnan_2022} and COSMOS-3D \citep{2024jwst.prop.5893K}. This includes the following filters: F090W, F115W, F182M, F200W, F210M, F277W, F335M, F356W, F410M, F444W, F460M, and F480M. Photometric counterparts are identified by matching to the spectroscopic targets within a 0.3 arcsec radius.

For SED fitting, we select targets whose available medium-resolution grating spectra can be combined to provide rest-frame spectral coverage over 3800--6700 \AA. Small instrumental detector gaps may still be present within this wavelength range. To remove low-quality edge regions, we exclude 20 pixels from each end of each spectrum, corresponding to a typical rest-frame width of about 60 \AA.

We then sequentially combine the available medium-resolution spectra, ordered by wavelength coverage. If an overlapping wavelength range exists, we estimate a multiplicative flux-normalization factor from the overlapping region. To do this, we first restrict the spectra to the overlapping wavelength region and sigma-clip them at the $2\sigma$ level. We then fit each overlap spectrum with a smooth polynomial and define the scaling factor as the median ratio of the two best-fit functions over the overlap region. The uncertainty in this scaling factor is estimated from the 16th and 84th percentiles of the ratio distribution.

We apply this normalization factor multiplicatively to the lower-wavelength spectrum. Its flux uncertainty is propagated by combining the original relative flux uncertainty with the relative uncertainty of the scale factor in quadrature. In the overlapping wavelength region, the scaled lower-wavelength spectrum is again evaluated on the wavelength grid of the higher-wavelength spectrum and combined with the higher-wavelength spectrum using inverse-variance weighting. The uncertainty of the combined overlap spectrum is computed from the sum of the inverse variance weights. The final stitched spectrum for each pair consists of the scaled lower-wavelength spectrum blue-ward of the overlap, the inverse-variance-weighted combined spectrum within the overlap, and the unscaled higher-wavelength spectrum red-ward of the overlap. For sources with more than two available medium-resolution gratings, this pairwise procedure is applied iteratively until a single stitched spectrum is obtained.

We use final stitched spectra spanning the rest-frame interval 3800--6700\,\AA\ for subsequent analysis, which covers key stellar continuum and interstellar features, including the Ca~H\&K doublet, the Balmer series (H$\delta$, H$\gamma$, H$\beta$, H$\alpha$), [O III] $\lambda\lambda$4959,5007, Mg~$b$, and Na\,\textsc{i}\,D doublet.

\subsection{Stellar Population Synthesis Modeling and Galaxy Classification}
\label{sec:sed}

\begin{figure*}[t]
    \centering
    \includegraphics[width=\textwidth]{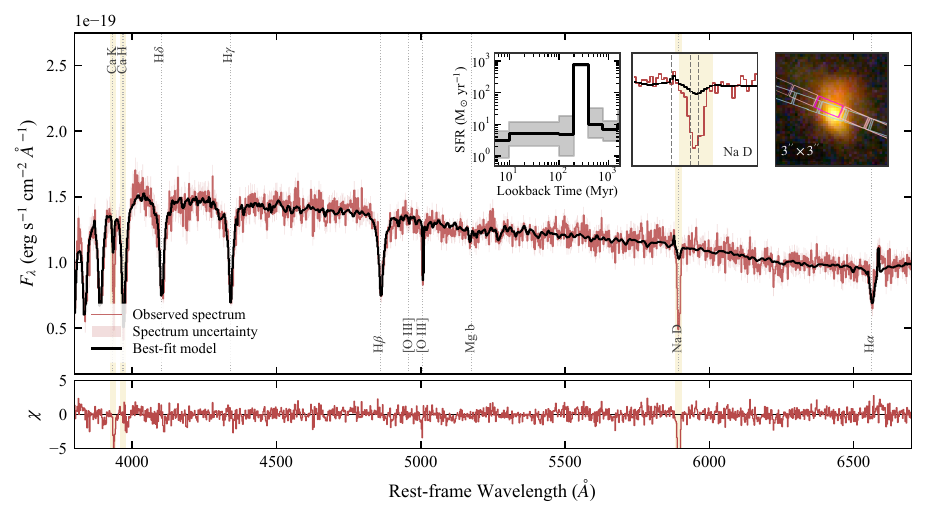}
    \caption{Representative SED fitting result for a quiescent galaxy at $z=3.71$. The broadband photometry is fit simultaneously with the spectrum but is not shown for clarity. The observed rest-frame optical spectrum is shown in red, with the best-fit model overlaid in black; the red shaded region denotes the $1\sigma$ spectral uncertainty. Vertical dotted lines indicate major spectral features, and the pale yellow shaded regions mark wavelength intervals masked during the SED fitting around Ca H\&K and Na\,\textsc{i}\,D. The inset panels show, from left to right, the inferred star formation history, a zoom-in around the Na\,\textsc{i}\,D absorption feature, and a $3'' \times 3''$ JWST/NIRCam RGB cutout constructed from F115W, F277W, and F444W imaging, with the NIRSpec slit position overlaid. The lower panel shows the residuals in units of the effective uncertainty. The fit reproduces the continuum shape and the main absorption features, while the inferred low recent star formation activity is consistent with a quiescent stellar population.}
    \label{fig:rep_sed}
\end{figure*}

To quantify neutral gas absorption features, it is necessary to first accurately model the stellar continuum contribution to the observed spectra, including the stellar Na\,\textsc{i}\,D absorption. We therefore perform SED fitting using \textsc{Prospector} (version 1.4.1; \citealt{Johnson_2021}), a Bayesian SED-fitting framework that enables simultaneous modeling of spectroscopic and photometric data. From this fitting, we infer physical properties including stellar mass ($M_\star$), star formation rate (SFR), metallicity, dust attenuation, and star formation history (SFH).

We adopt FSPS (Flexible Stellar Population Synthesis; \citealt{Conroy_2009, Conroy_2010}) for stellar population synthesis, using the MIST isochrones \citep{Choi_2016}, the MILES spectral library, and assuming a Chabrier IMF \citep{Chabrier_2003}, following a similar setup as in \citet{Setton_2023} and 
\citet{khullar2026leggosijwstleggos}. We model the SFH using a nonparametric approach with seven time bins, with bin edges defined as a function of the age of the Universe at the galaxy redshift. A continuity prior is applied to the logarithmic SFR ratios between adjacent bins, modeled as a Student-$t$ distribution \citep{Leja_2019}. Dust attenuation is modeled using a two-component model \citep{Charlot_2000} adopting the \citet{Kriek_2013} attenuation law, with free parameters including the $V$-band optical depth, the power-law slope of the attenuation curve, and the multiplicative factor for additional attenuation toward young stars. To model nebular emission accurately, we adopt the nebular marginalization approach \citep{Johnson_2021}, which fits emission-line fluxes directly via linear least squares at each likelihood evaluation. 

We fit the spectra and photometry simultaneously. To account for normalization offsets between the spectra and photometry, including possible slit-loss or flux-calibration effects, we apply a multiplicative spectral calibration vector modeled as a fifth-order Chebyshev polynomial and include a spectral jitter term to account for potential underestimation of spectroscopic uncertainties. We mask the Na\,\textsc{i}\,D ($\lambda\lambda5891,5897$\,\AA) and Ca H\&K ($\lambda\lambda3934,3968$\,\AA) regions because they can contain both stellar and ISM absorption components, then use the Markov chain Monte Carlo (MCMC) sampler \texttt{emcee} \citep{Foreman_Mackey_2013} to sample the posterior. The final physical properties are inferred from the posterior distributions, with the 16th, 50th, and 84th percentiles adopted as the lower, median, and upper uncertainties, respectively. Figure~\ref{fig:rep_sed} shows a representative SED fitting result for a quiescent galaxy at $z=3.71$, illustrating the observed spectrum, best-fit stellar continuum model, masked wavelength intervals, inferred SFH, and residual spectrum used for the subsequent absorption-line analysis.

We then subtract the best-fit stellar continuum model from the observed spectrum to construct residual spectra. These residuals trace interstellar absorption features associated with neutral gas; in particular, since emission lines are modeled and subtracted as part of the nebular marginalization procedure, residuals around the Na\,\textsc{i}\,D region directly trace neutral gas absorption without contamination from nebular emission. Moreover, since our stellar continuum model includes the Na\,\textsc{i}\,D absorption intrinsic to the underlying stellar populations, this stellar-origin component is removed, ensuring that the residual Na\,\textsc{i}\,D signal reflects interstellar rather than stellar absorption \citep{Baron_2021,Davies_2024,taylor2026jwstexcelssurveyoutflows}.

\begin{figure}[t]
    \centering
    \includegraphics[width=\columnwidth]{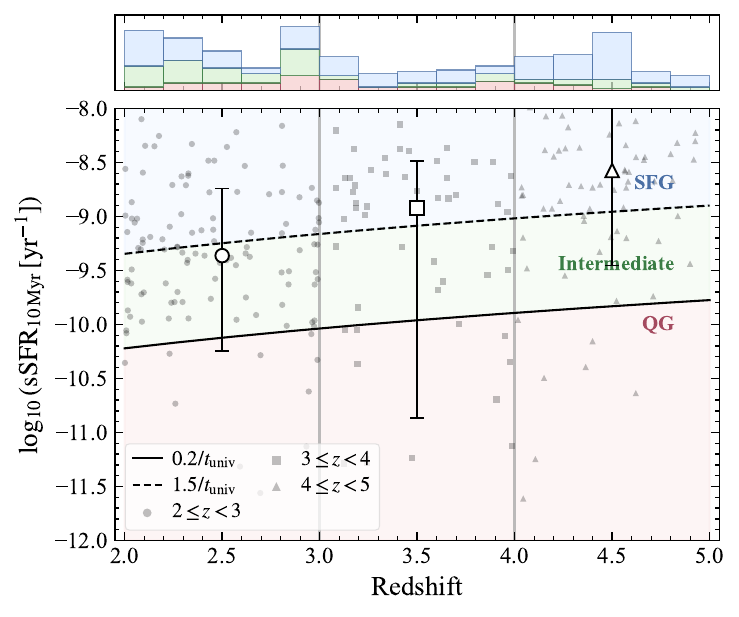}
    \caption{Classification of galaxies by specific star formation rate. We show $\log_{10}({\rm sSFR}_{10\,{\rm Myr}})$ as a function of redshift for the final SED-fitting sample at $2 \leq z < 5$. Gray circles, squares, and triangles denote individual galaxies in the redshift intervals $2 \leq z < 3$, $3 \leq z < 4$, and $4 \leq z < 5$, respectively. Large open symbols mark the median $\log_{10}({\rm sSFR}_{10\,{\rm Myr}})$ in each redshift bin, plotted at the bin center, and the vertical error bars show the 16th--84th percentile range of the distribution. The top histogram shows the redshift distribution of the sample. The solid and dashed curves correspond to $0.2/t_{\rm univ} (z)$ and $1.5/t_{\rm univ} (z)$, which separate the quiescent, intermediate, and star-forming regimes shown by the shaded regions. This diagnostic identifies galaxies with suppressed recent star formation relative to the cosmic-time-scaled thresholds, including a quiescent subset below $0.2/t_{\rm univ}(z)$.}
    \label{fig:ssfrvsz}
\end{figure}

We classify galaxies into three populations based on their specific star formation rate (sSFR), calculated as $\mathrm{SFR}/M_\star$, with the SFR derived from SED fitting and averaged over a 10Myr timescale. The classification thresholds are defined relative to the age of the Universe at the redshift of each galaxy, $t_{\rm univ} (z)$. Galaxies with $\mathrm{sSFR} \leq 0.2\,t_{\rm univ}(z)^{-1}$ are classified as quiescent, those with $\mathrm{sSFR} \geq 1.5\,t_{\rm univ}(z)^{-1}$ as star-forming, and those with sSFR values in between as the intermediate population \citep{Carnall_2020,Tacchella_2022}. This selection yields an initial sample of 308 galaxies. After removing 34 galaxies with poorly normalized spectra, as described in the next section, the final sample used for the subsequent analysis contains 274 galaxies. The distribution of this final sample is shown in Figure~\ref{fig:ssfrvsz}.

\subsection{Spectral Stacking}
\label{sec:stacking}

We construct rest-frame stacked residual spectra in the Na\,\textsc{i}\,D region from one-dimensional spectroscopic observations of individual galaxies. All spectra are first shifted to the rest frame using their spectroscopically confirmed redshifts. Then, for each absorption feature, we define a 300\,\AA-wide rest-frame window centered on the absorption-line doublet. 
To avoid artificial oversampling during spectral stacking, we group spectra into redshift bins with different wavelength-pixel spacings: $\Delta\lambda = 0.90$, $0.70$, and $0.60$\,\AA\ for $2\leq z<3$, $3\leq z<4$, and $4\leq z <5$, respectively. Within each group, a common rest-frame wavelength grid is constructed, and all spectra are resampled using \textsc{SpectRes} \citep{carnall2017spectresfastspectralresampling}, which ensures flux conservation and consistent uncertainty propagation.

We normalize each residual spectrum in the continuum by fitting a fifth-order polynomial to the spectral regions outside the Na\,\textsc{i}\,D absorption region. To mitigate noise, we first apply a one-dimensional median filter with a kernel width of 11 pixels before fitting. The normalized residual spectrum is then
\begin{equation}
F_{\mathrm{res}}(\lambda) = F_{\mathrm{obs}}(\lambda) - F_{\mathrm{model}}(\lambda),
\end{equation}
\begin{equation}
    F_{\mathrm{norm}}(\lambda) = \frac{F_{\mathrm{res}}(\lambda)}{C(\lambda)},
\end{equation}
where $C(\lambda)$ is the fitted continuum model.

To quantify the normalization quality, we measure the mean squared deviation of the normalized flux from zero in the continuum regions,
\begin{equation}
    \mathrm{MSE} = \left\langle \left(F_{\mathrm{norm}}\right)^2 \right\rangle,
\end{equation}
and assign each spectrum a weight $w = 1/\mathrm{MSE}$. Spectra with $w < 1$ are excluded from the stack. The final sample sizes per redshift bin and galaxy type are summarized in Table~\ref{table:stack}.

\begin{deluxetable}{lccc}
\tablewidth{0pt}
\tablecaption{Number of stacked targets}
\label{table:stack}
\tablehead{
\colhead{Type} & \colhead{$z=2$--3} & \colhead{$z=3$--4} & \colhead{$z=4$--5}
}
\startdata
Star-forming & 55 & 34 & 62 \\
Intermediate & 50 & 10 & 14 \\
Quiescent & 22 & 17 & 10
\enddata
\tablecomments{Columns indicate the number of galaxies in each redshift bin ($z=2$--3, $3$--4, and $4$--5) classified by spectral type (star-forming, intermediate, and quiescent; see Section~\ref{sec:sed}). Sample sizes correspond to the final stacked sample after excluding spectra with weight $w < 1$ (Section~\ref{sec:stacking}), reducing the total from 308 to 274 galaxies.}
\end{deluxetable}

We stack the normalized spectra using a weighted median, which is robust to outliers and non-Gaussian flux distributions. At each wavelength pixel $\lambda_j$, the normalized flux values $\{f_1, f_2, \ldots, f_n\}$ are sorted in ascending order with their associated weights $\{w_1, w_2, \ldots, w_n\}$. The weighted median is defined as the value $f_k$ satisfying
\begin{equation}
\frac{\sum_{i=1}^{k-1} w_i}{\sum_{i=1}^{n} w_i} \leq \frac{1}{2} \quad \text{and} \quad \frac{\sum_{i=k+1}^{n} w_i}{\sum_{i=1}^{n} w_i} \leq \frac{1}{2}.
\end{equation}

Uncertainties for each stacked spectrum are estimated via bootstrap resampling: within each group, we draw a sample of equal size with replacement, compute the weighted-median stack, and repeat this process 1000 times. The standard deviation of the bootstrap distribution at each wavelength pixel is adopted as the $1\sigma$ uncertainty.

\subsection{Absorption Line Fitting}
\label{sec:fitting}

\begin{figure*}[t]
    \centering
    \includegraphics[width=\textwidth]{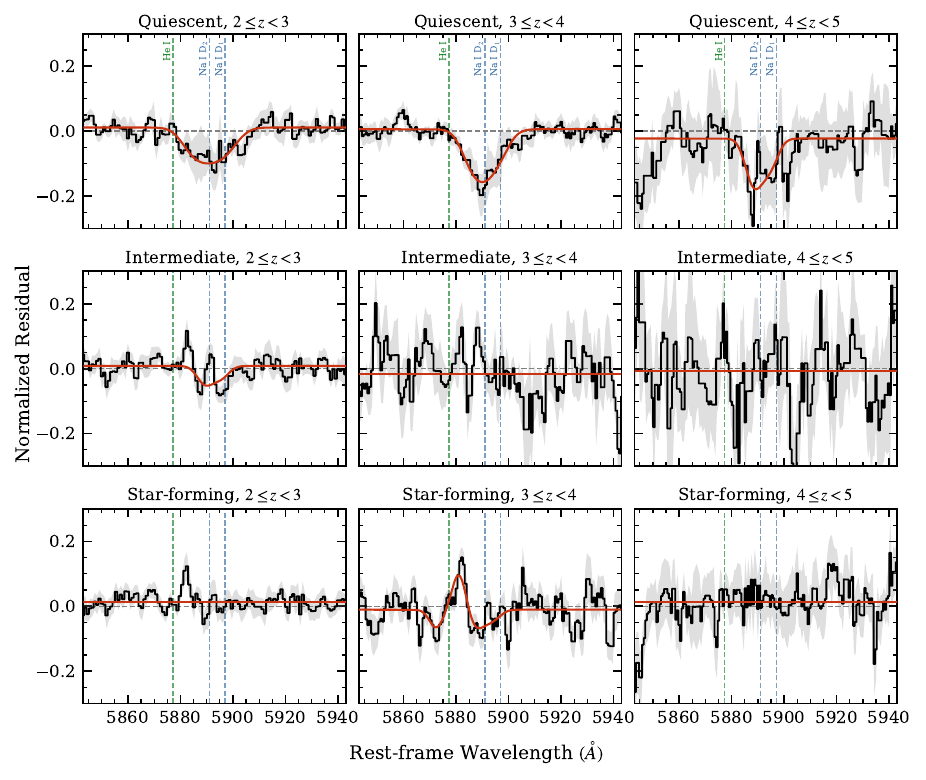}
    \caption{Model fits to the stacked residual spectra around the Na\,\textsc{i}\,D region. The panels are arranged by sSFR and redshift, with rows showing quiescent, intermediate, and star-forming galaxies and columns showing $2 \leq z < 3$, $3 \leq z < 4$, and $4 \leq z < 5$. Black step curves show the weighted-median stacked residual spectra, and gray shaded regions indicate the bootstrap uncertainties. Red curves show the best-fit spectral models selected using the Bayesian information criterion. The models include a local continuum and, where favored, Na\,\textsc{i}\,D absorption together with He\,\textsc{i} emission and/or He\,\textsc{i} absorption components constrained relative to the systemic velocity measured from the H$\alpha$--[N\,\textsc{ii}] stack. The horizontal dashed gray line marks zero residual, while the vertical dashed lines indicate the expected rest-frame wavelengths of He\,\textsc{i} $\lambda5877$ and the Na\,\textsc{i}\,D doublet at $\lambda5891$ and $\lambda5897$. The number of galaxies contributing to each stack is listed in Table~\ref{table:stack}. Quiescent stacks show negative residuals near the Na\,\textsc{i}\,D doublet, most prominently at $3 \leq z < 4$, while the intermediate and star-forming stacks show weaker or less coherent features.}
    \label{fig:fit}
\end{figure*}

A key complication in fitting the Na\,\textsc{i}\,D doublet is contamination from the He\,\textsc{i} $\lambda5877$ emission line, which lies just blueward of Na\,\textsc{i}\,D $\lambda5891$ and can partially fill in or distort the blue wing of the absorption profile. Moreover, He\,\textsc{i} can exhibit a P-Cygni profile \citep{Guseva_2023}, a combination of nebular emission at the systemic velocity and blueshifted absorption, in galaxies with strong winds. If unaccounted for, such a blueshifted He\,\textsc{i} absorption component could mimic or enhance the apparent Na\,\textsc{i}\,D outflow signal. We therefore adopt a model comparison framework to disentangle these components rather than assuming a fixed model structure.
 
We fit six nested models for each combination of galaxy type (star-forming, intermediate, and quiescent) and redshift bin ($2\leq z <3$, $3\leq z <4$, and $4\leq z<5$) within a 150\AA\,\,rest-frame window centered on the Na\,\textsc{i}\,D doublet, and select the best-fitting model using the Bayesian Information Criterion (BIC). The six models are: (1) linear continuum only, (2) continuum with He\,\textsc{i} emission, (3) continuum with He\,\textsc{i} P-Cygni feature, (4) continuum with Na\,\textsc{i}\,D absorption, (5) continuum with Na\,\textsc{i}\,D absorption and He\,\textsc{i} emission, and (6) continuum with Na\,\textsc{i}\,D absorption and He\,\textsc{i} P-Cygni feature. We require the selected model to have the lowest BIC and to be preferred over the next-best model by $\Delta{\rm BIC}>10$ \citep{Liddle_2007}. We then perform MCMC sampling only for the selected model.
 
The Na\,\textsc{i}\,D absorption is parametrized using the partial covering model of \citet{Rupke_2005a}:
\begin{equation}
F_{\mathrm{Na\,\textsc{i}\,D}}(\lambda)
=
C_f
\left[
\exp\left(
-\tau_b(\lambda)-\tau_r(\lambda)
\right)-1
\right],
\end{equation}
where $C_f$ is the covering fraction of the absorbing gas, and $\tau_b(\lambda)$ and $\tau_r(\lambda)$ are the optical depth profiles of the blue (Na\,\textsc{i}\,D $\lambda5891$) and red (Na\,\textsc{i}\,D $\lambda5897$) components, respectively. Each optical depth profile follows a Gaussian in velocity space:
\begin{equation}
\tau(v)
=
\tau_0
\exp\left(
-\frac{v^2}{2\sigma_{\mathrm{NaD}}^2}
\right).
\end{equation}
The central optical depth of the blue component is fixed to twice that of the red component, $\tau_{0,b} = 2\tau_{0,r}$, reflecting the known oscillator strength ratio of the two transitions. Both components share a common velocity offset $v_{\mathrm{NaD}}$ and velocity dispersion $\sigma_{\mathrm{NaD}}$.
 
The He\,\textsc{i} emission component is modeled as a Gaussian anchored to the systemic velocity $v_\mathrm{sys}$ and velocity dispersion $\sigma_\mathrm{sys}$ measured independently from the H$\alpha$ and [NII] stacked spectrum of the same galaxy population and redshift bin. The He\,\textsc{i} emission velocity is constrained to vary within a limited range around $v_\mathrm{sys}$, and its velocity dispersion is similarly constrained near $\sigma_\mathrm{sys}$. For the models that include the P-Cygni feature, an additional He\,\textsc{i} absorption component is included with its velocity restricted to $v < v_\mathrm{sys}$.
 
We assign uniform priors over physically motivated ranges to all free parameters and initialize the MCMC walkers near the least-squares solution. Parameter inference is performed using \texttt{emcee} \citep{Foreman_Mackey_2013}. For reproducibility, the full set of free parameters, prior ranges, and initialization procedure are listed in Table~\ref{tab:naid_priors} in Appendix~\ref{app:naid_priors}.
 
An outflow is identified when $v_{84} < 0$\,km\,s$^{-1}$ and a robust outflow requires $v_{50} \leq -100$\,km\,s$^{-1}$. The characteristic outflow velocity is defined as
\begin{equation}
v_{\mathrm{out}} = |v_{\mathrm{NaD}}| + 2\sigma_{\mathrm{NaD}},
\end{equation}
which captures the maximum extent of the blue wing of the absorption profile.

\subsection{Outflow Properties}
\label{sec:outflow-props}

\begin{deluxetable*}{lllccccccccc}
\tabletypesize{\scriptsize}
\tablewidth{0pt}
\tablecaption{Derived Outflow Properties from Na\,\textsc{i}\,D Stacking and Fitting}
\tablehead{
\colhead{Type} &
\colhead{$z$} &
\colhead{$N$} &
\colhead{$\log M_\star$} &
\colhead{$\log {\rm SFR}$} &
\colhead{$|v_{\rm Na\,\textsc{i}\,D}|$} &
\colhead{$\sigma_{\rm Na\,\textsc{i}\,D}$} &
\colhead{$v_{\rm out}$} &
\colhead{$\log M_{\rm neutral}$} &
\colhead{$\log \dot{M}_{\rm out}$} &
\colhead{$\log \dot{E}_{\rm out}$} &
\colhead{$\log \eta$} \\
\colhead{} &
\colhead{} &
\colhead{} &
\colhead{[$M_\odot$]} &
\colhead{[$M_\odot\,{\rm yr}^{-1}$]} &
\colhead{[km s$^{-1}$]} &
\colhead{[km s$^{-1}$]} &
\colhead{[km s$^{-1}$]} &
\colhead{[$M_\odot$]} &
\colhead{[$M_\odot\,{\rm yr}^{-1}$]} &
\colhead{[erg s$^{-1}$]} &
\colhead{}
}
\startdata
\\
SFG & 2--3 & 55 & $9.93^{+0.61}_{-0.52}$ & $0.95^{+0.45}_{-0.42}$ & \nodata & \nodata & \nodata & \nodata & \nodata & \nodata & \nodata \\[4pt]
SFG & 3--4 & 34 & $9.94^{+0.58}_{-0.46}$ & $1.21^{+0.51}_{-0.57}$ & $164^{+66}_{-49}$ & $79^{+30}_{-21}$ & $324^{+103}_{-68}$ & $6.95^{+0.13}_{-0.11}$ & $0.46^{+0.20}_{-0.15}$ & $41.0^{+0.4}_{-0.3}$ & $-0.69^{+0.70}_{-0.71}$ \\[4pt]
SFG & 4--5 & 62 & $9.23^{+0.61}_{-0.49}$ & $0.81^{+0.45}_{-0.49}$ & \nodata & \nodata & \nodata & \nodata & \nodata & \nodata & \nodata \\[4pt]
INT & 2--3 & 50 & $10.24^{+0.85}_{-0.46}$ & $0.47^{+0.71}_{-0.52}$ & $79^{+22}_{-21}$ & $75^{+24}_{-19}$ & $230^{+49}_{-41}$ & $7.05^{+0.15}_{-0.14}$ & $0.42^{+0.17}_{-0.16}$ & $40.7^{+0.3}_{-0.3}$ & $-0.09^{+0.75}_{-0.90}$ \\[4pt]
INT & 3--4 & 10 & $10.18^{+0.44}_{-0.43}$ & $0.49^{+0.52}_{-0.47}$ & \nodata & \nodata & \nodata & \nodata & \nodata & \nodata & \nodata \\[4pt]
INT & 4--5 & 14 & $9.97^{+0.42}_{-0.28}$ & $0.60^{+0.55}_{-0.24}$ & \nodata & \nodata & \nodata & \nodata & \nodata & \nodata & \nodata \\[4pt]
QG & 2--3 & 22 & $10.75^{+0.49}_{-0.48}$ & $-1.69^{+1.73}_{-3.90}$ & $105^{+34}_{-24}$ & $232^{+40}_{-34}$ & $574^{+77}_{-67}$ & $7.70^{+0.16}_{-0.14}$ & $1.47^{+0.15}_{-0.13}$ & $42.5^{+0.2}_{-0.2}$ & $3.45^{+5.99}_{-2.04}$ \\[4pt]
QG & 3--4 & 17 & $11.05^{+0.32}_{-0.37}$ & $-0.42^{+1.16}_{-2.92}$ & $126^{+29}_{-27}$ & $189^{+31}_{-29}$ & $505^{+54}_{-49}$ & $7.64^{+0.12}_{-0.09}$ & $1.36^{+0.10}_{-0.09}$ & $42.3^{+0.1}_{-0.1}$ & $1.68^{+1.64}_{-1.18}$ \\[4pt]
QG & 4--5 & 10 & $10.62^{+0.36}_{-0.45}$ & $-0.52^{+0.76}_{-3.14}$ & $128^{+39}_{-32}$ & $89^{+32}_{-24}$ & $313^{+64}_{-53}$ & $7.43^{+0.13}_{-0.11}$ & $0.94^{+0.16}_{-0.15}$ & $41.4^{+0.3}_{-0.3}$ & $1.45^{+4.21}_{-0.74}$ \\[4pt]
\enddata
\tablecomments{Columns are as follows: galaxy type (SFG = star-forming, INT = intermediate, QG = quiescent), redshift bin, number of stacked galaxies ($N$), median stellar mass, median star formation rate, Na\,\textsc{i}\,D absorption velocity centroid, Na\,\textsc{i}\,D velocity dispersion, outflow velocity, neutral gas mass, mass outflow rate, kinetic energy outflow rate, and mass loading factor ($\eta \equiv \dot{M}_{\rm out}/{\rm SFR}$). Outflow properties are derived from Na\,\textsc{i}\,D absorption-line stacking and fitting for each galaxy population and redshift bin. The outflow velocity is defined as $v_{\rm out}=|v_{\rm NaD}|+2\sigma_{\rm NaD}$. Larger values of $|v_{\rm NaD}|$ indicate stronger (more blueshifted) outflows. Quoted uncertainties indicate the 16th, 50th, and 84th percentiles of the posterior distributions. Entries marked as \nodata\ correspond to Na\,\textsc{i}\,D non-detections, for which the outflow properties could not be robustly constrained.}
\label{table:props}
\end{deluxetable*}

From the Na\,\textsc{i}\,D absorption line fitting, we derive the neutral gas column density, mass outflow rate, and mass loading factor following \citet{Davies_2024} and \citet{Rupke_2005a}.

The Na I column density is computed from $\tau_{0,r}$ using the relation from \citet{2011piim.book.....D}:
\begin{equation}
\begin{aligned}
N(\mathrm{NaI})
&=
10^{13}\,\mathrm{cm}^{-2}
\left(
\frac{\tau_{0,r}}{0.7580}
\right)
\left(
\frac{0.4164}{f_{lu}}
\right) \\
&\quad\times
\left(
\frac{1215\,\text{\AA}}{\lambda_{lu}}
\right)
\left(
\frac{b}{10\,\mathrm{km\,s}^{-1}}
\right),
\end{aligned}
\end{equation}
where $f_{lu} = 0.32$ and $\lambda_{lu} = 5897$\,\AA\ are the oscillator strength and rest-frame wavelength of the red transition, and $b = \sqrt{2}\,\sigma_{\mathrm{NaD}}$ is the Doppler parameter. The neutral hydrogen column density $N(\mathrm{HI})$ is then obtained assuming Milky-Way-like sodium abundance and dust depletion factors \citep{Rupke_2005a} with a 10\% neutral fraction:
\begin{equation}
N(\mathrm{HI})
=
\frac{N(\mathrm{NaI})}{1-y}
10^{-(\log A_\mathrm{Na}+\delta_\mathrm{Na})},
\end{equation}
where $y = 0.9$ is the ionization fraction, $\log A_\mathrm{Na} = -5.69$ is the solar sodium abundance, and $\delta_\mathrm{Na} = -0.95$ is the dust depletion factor.

The total mass of neutral gas entrained in the outflow is 
\begin{equation}
\begin{split}
M_\mathrm{neutral}
\,(\mathrm{M}_\odot)
&=
5.6\times10^{7}
\left(\frac{C_\Omega}{0.4}\right)
C_f \\
&\quad\times
\left(\frac{N(\mathrm{HI})}{10^{21}\,\mathrm{cm}^{-2}}\right)
\left(\frac{r_\mathrm{out}}{1\,\mathrm{kpc}}\right)^{2}.
\end{split}
\label{eq:mneutral}
\end{equation}
following \citet{Rupke_2005a}

The neutral gas mass outflow rate is estimated using the time-averaged shell model of \citet{Rupke_2005a}, as updated in \citet{Baron_2021}:
\begin{equation}
\begin{split}
\dot{M}_\mathrm{out}
\,(\mathrm{M}_\odot\,\mathrm{yr}^{-1})
&=
11.45
\left(\frac{C_\Omega}{0.4}\right)
C_f \\
&\quad\times
\left(\frac{N(\mathrm{HI})}{10^{21}\,\mathrm{cm}^{-2}}\right)
\left(\frac{r_\mathrm{out}}{1\,\mathrm{kpc}}\right) \\
&\quad\times
\left(\frac{v_\mathrm{out}}{200\,\mathrm{km\,s}^{-1}}\right).
\end{split}
\label{eq:mdot}
\end{equation}
where $C_\Omega = 0.5$ is the large-scale covering factor of the wind \citep{Rupke_2005b}, and $r_\mathrm{out} = 1$\,kpc is the assumed outflow radius \citep{Davies_2024}.

The kinetic energy outflow rate is derived from the mass outflow rate as $\dot{E}_\mathrm{out} = \frac{1}{2}\dot{M}_\mathrm{out}v_\mathrm{out}^2$ \citep{Rupke_2005a,Baron_2021}, which can be expressed as
\begin{equation}
\begin{split}
\dot{E}_\mathrm{out}\,(\mathrm{erg\,s}^{-1})
&=
1.47\times10^{41}
\left(\frac{C_\Omega}{0.4}\right)
C_f \\
&\quad\times
\left(\frac{N(\mathrm{HI})}{10^{21}\,\mathrm{cm}^{-2}}\right)
\left(\frac{r_\mathrm{out}}{1\,\mathrm{kpc}}\right) \\
&\quad\times
\left(\frac{v_\mathrm{out}}{200\,\mathrm{km\,s}^{-1}}\right)^{3}.
\end{split}
\end{equation}

The mass loading factor is defined as
\begin{equation}
\eta = \dot{M}_\mathrm{out}/\mathrm{SFR},
\end{equation}

where the SFR is taken from the \textsc{Prospector} SPS posterior (Section~\ref{sec:sed}). All quantities are computed using the full MCMC posterior distributions, and the reported values correspond to the 16th, 50th, and 84th percentiles.

\begin{figure*}[t]
    \centering
    \includegraphics[width=\textwidth]{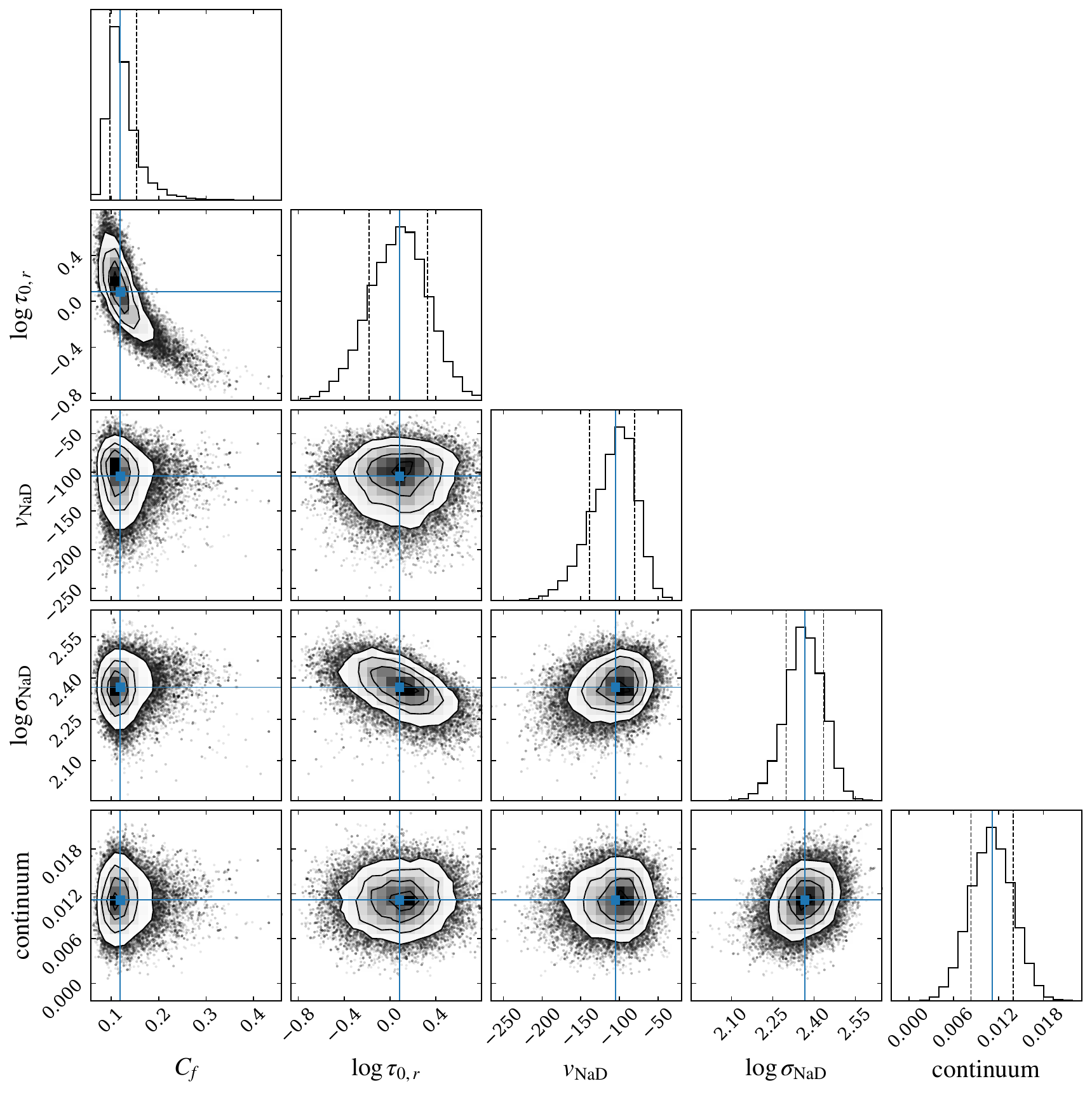}
    \caption{Posterior corner plot for the best-fit Na\,\textsc{i}\,D absorption model. Diagonal panels show the marginalized one-dimensional posteriors, and off-diagonal panels show the two-dimensional joint posteriors for $C_f$, $\log \tau_{0,r}$, $v_{\rm NaD}$, $\log \sigma_{\rm NaD}$, and continuum. Black points are posterior samples, contours enclose the highest-density credible regions, blue lines indicate the posterior medians, and dashed lines mark the 16th and 84th percentiles. The posterior structure shows a strong covariance between covering fraction and optical depth, while the Na\,\textsc{i}\,D velocity and dispersion remain relatively well constrained.}
    \label{fig:corner}
\end{figure*}

\begin{figure}[t]
    \centering
    \includegraphics[width=\columnwidth]{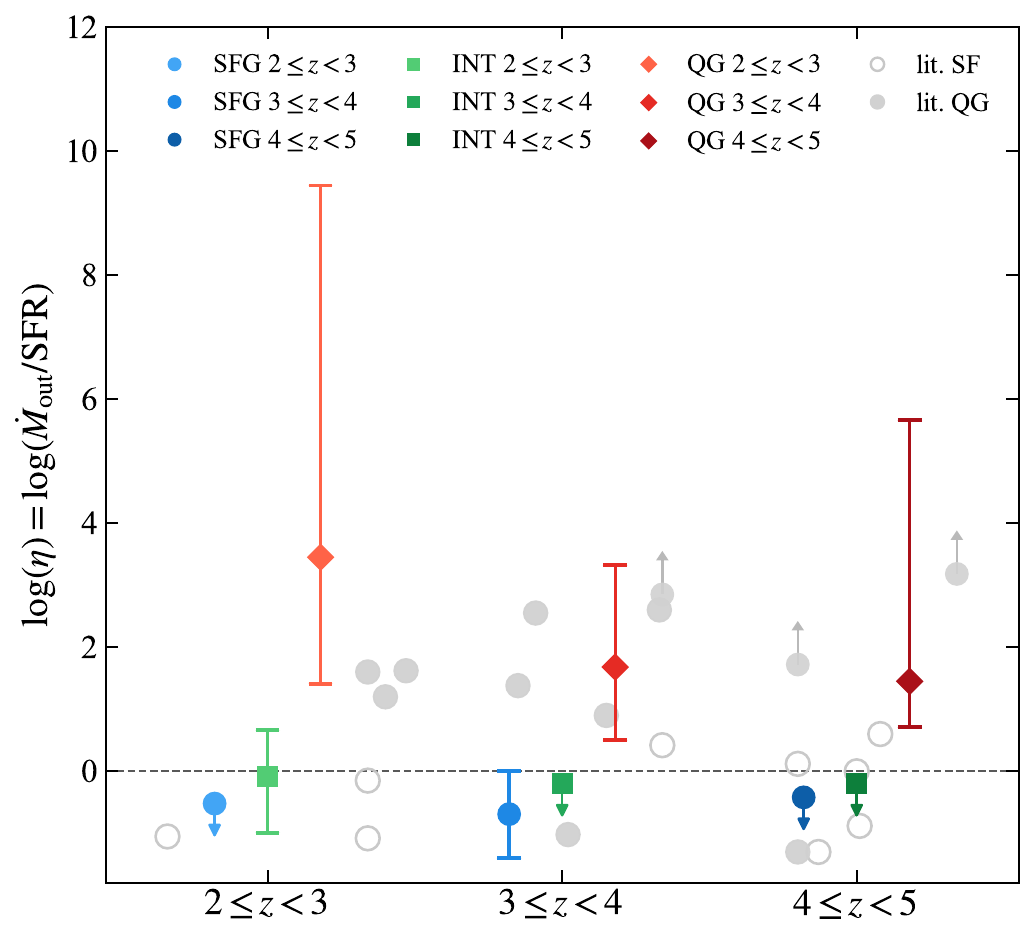}
    \caption{Comparison of the neutral gas mass loading factor, $\eta=\dot{M}_{\rm out}/{\rm SFR}$, for our stacked galaxy samples and literature neutral gas outflow samples with available $\eta$ or SFR-based estimates. Colored symbols show the star-forming, intermediate, and quiescent stacks in this work, with symbol shape indicating galaxy type and color shade indicating redshift bin; error bars show the 16th--84th percentile ranges. Downward arrows mark qualitative upper limits for stacks without significant Na\,\textsc{i}\,D outflow detections, estimated by combining the precise stacked SFR of each non-detected bin with the characteristic $\dot{M}{\rm out}$ inferred from the detected stack of the same galaxy class. Gray symbols show literature measurements of neutral gas outflows separated into star-forming and quiescent systems. Literature points represent individual galaxies or small samples at similar redshifts, whereas the points from this work represent stacked measurements for uniformly classified galaxy populations in fixed redshift bins. The literature comparison includes the Blue Jay Na\,\textsc{i}\,D outflows from \citet{Davies_2024}, recently quenched galaxies from \citet{Valentino_2025}, quiescent galaxies from \citet{taylor2026jwstexcelssurveyoutflows} and \citet{sun2026censusnadtracedneutral}. The dashed horizontal line marks $\eta=1$. The quiescent stacks show elevated mass loading factors relative to the star-forming and intermediate stacks in this work and are comparable to quiescent literature systems, suggesting that their neutral gas outflows are unusually efficient relative to their current star formation.}
    \label{fig:eta_comparison}
\end{figure}

\begin{figure*}[t]
    \centering
    \includegraphics[width=\textwidth]{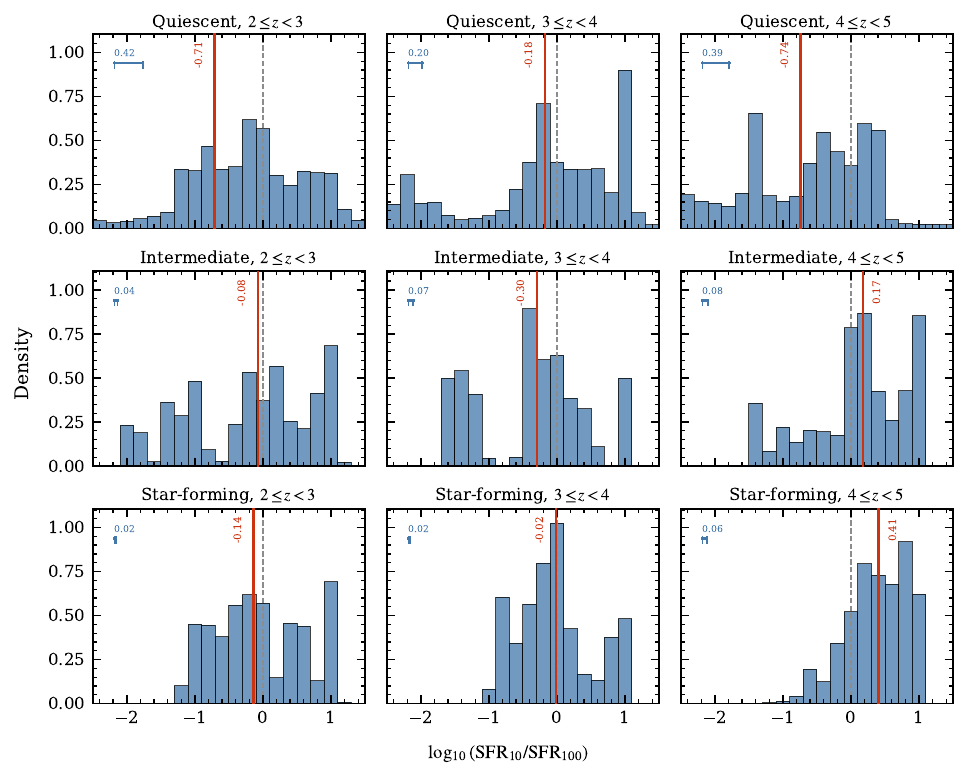}
    \caption{Distribution of the recent-to-past star formation rate ratio, $\log_{10}(\mathrm{SFR}_{10}/\mathrm{SFR}_{100})$, for each galaxy population and redshift bin. Here, $\mathrm{SFR}_{10}$ and $\mathrm{SFR}_{100}$ denote the star formation rates averaged over the most recent 10 Myr and 100 Myr, respectively. In each panel, the histogram is constructed from Monte Carlo realizations that propagate their individual posterior uncertainties on $\mathrm{SFR}_{10}$ and $\mathrm{SFR}_{100}$. The red vertical line marks the pooled median, with its value annotated in red; the blue horizontal bar indicates the median $1\sigma$ uncertainty on the ratio. The gray dashed vertical line marks zero, corresponding to a constant star formation rate over the past 100\,Myr. The negative medians in all three quiescent bins indicate declining recent star formation on short timescales.}
    \label{fig:sfr_ratio}
\end{figure*}

\begin{figure}[t]
    \centering
    \includegraphics[width=\columnwidth]{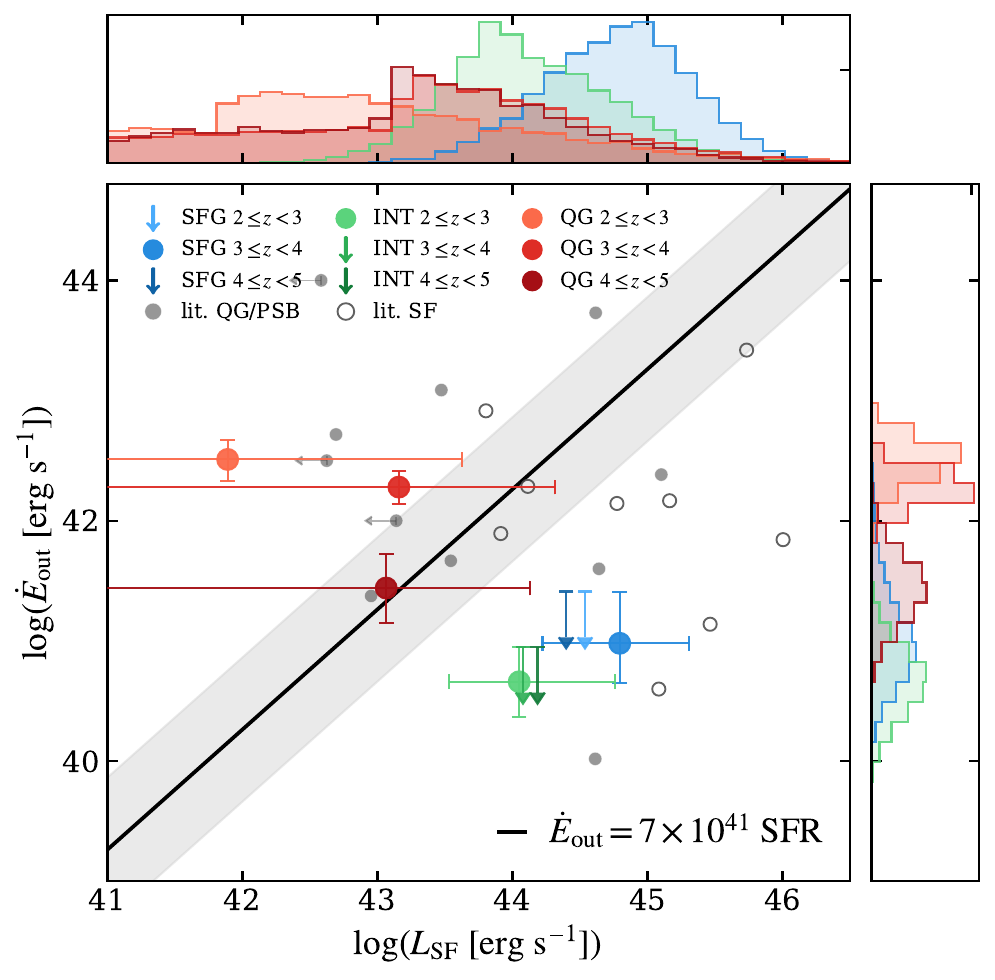}
    \caption{Comparison between the neutral gas outflow kinetic energy rate, $\dot{E}_{\rm out}$, and the luminosity associated with star formation, $L_{\rm SF}$, for our stacked galaxy samples. Blue, green, and red symbols denote the SFG, intermediate, and quiescent galaxy samples, respectively, with darker shades indicating higher redshift within each class. Downward arrows mark qualitative upper limits for bins without direct $\dot{E}_{\rm out}$ measurements; the arrow lengths are for visualization only. Small gray markers show literature comparison samples, with filled circles denoting quiescent or post-starburst systems and open circles denoting star-forming systems, compiled from \citet{Baron_2021,Davies_2024,Valentino_2025,taylor2026jwstexcelssurveyoutflows}, and \citet{sun2026censusnadtracedneutral}. The top and right panels show the marginalized uncertainty distributions of $L_{\rm SF}$ and $\dot{E}_{\rm out}$, respectively, obtained by combining the Na\,\textsc{i}\,D fitting posterior chains with the uncertainties in the stacked SFRs. The black solid line shows the expected mechanical energy injection rate from supernova feedback, $\dot{E}_{\rm out}=7\times10^{41}{\rm SFR}$, following the Starburst99-based calibration used by \citet{Veilleux_2005}. The gray shaded region indicates a representative systematic uncertainty of 0.6 dex in the inferred outflow energy rate. The quiescent galaxy stacks lie above the energy expected from their current star formation, suggesting that instantaneous star formation alone may not fully account for the observed energetics of neutral gas outflows.}
    \label{fig:EoutvsLSF}
\end{figure}

\begin{figure*}[t]
    \centering
    \includegraphics[width=\textwidth]{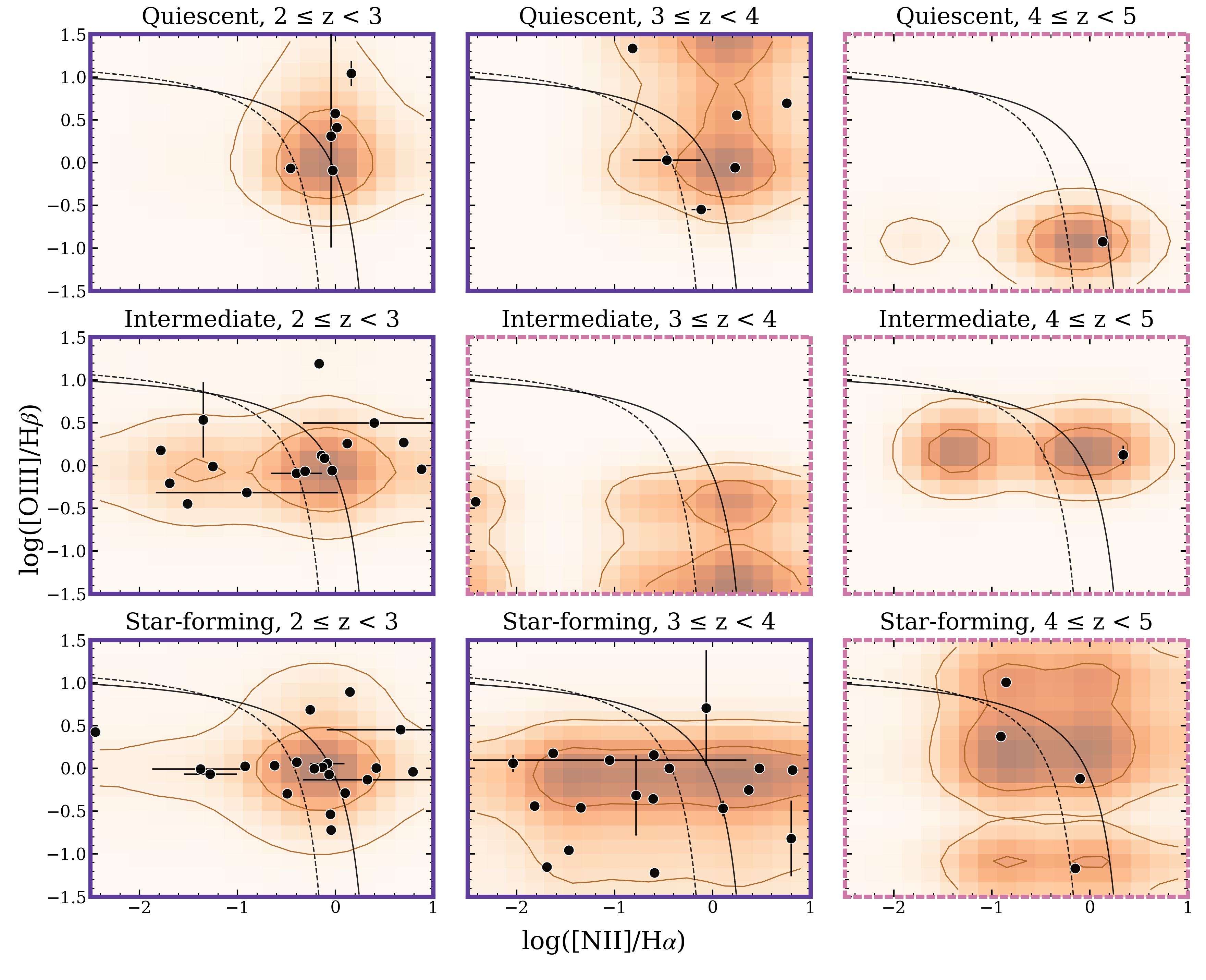}
    \caption{BPT joint distribution of $\log([\rm{N}\,\textsc{ii}]/{\rm H}\alpha)$ and $\log([\rm{O}\,\textsc{iii}]/{\rm H}\beta)$ for each galaxy population and redshift bin. The orange shaded background and contours show the hypothetical joint distribution constructed by Monte Carlo resampling of the individual posterior distributions of each line ratio, representing all possible pairings of $[\rm{N}\,\textsc{ii}]/{\rm H}\alpha$ and $[\rm{O}\,\textsc{iii}]/{\rm H}\beta$ measurements within each bin. Black points with error bars indicate individual galaxies for which both line ratios are simultaneously detected, with error bars representing the $1\sigma$ posterior uncertainties. The solid and dashed curves show the maximum-starburst line of \citet{Kewley_2001} and the composite demarcation of \citet{Kauffmann_2003}, respectively. Panels outlined with solid purple borders contain six or more galaxies with jointly detected line ratios and are considered statistically reliable; panels outlined with pink dashed borders contain fewer than five such galaxies and should be interpreted with caution. In the statistically reliable quiescent bins at $2 \leq z<3$ and $3 \leq z<4$, the joint distributions extend toward the AGN locus, and individual detections are preferentially located above the Kauffmann demarcation line, consistent with a possible AGN activity contribution.}
    \label{fig:bpt_2d_histo}
\end{figure*}

\section{Results} \label{sec:results}

\subsection{Detection of Neutral Gas Outflows}

Na\,\textsc{i}\,D absorption is detected in all three quiescent galaxy bins, while only one bin each in the intermediate and star-forming populations shows a significant detection. The consistent detection across all redshift bins in the quiescent population is particularly notable given their relatively small sample sizes.

Figure~\ref{fig:fit} shows the stacked residual spectra following continuum subtraction, together with the absorption-line models. The selected models reproduce the detected Na\,\textsc{i}\,D profiles well, including the consistently blueshifted absorption seen in the quiescent stacks and the weaker or absent absorption in the intermediate and star-forming stacks.

Model selection via BIC favors a pure Na\,\textsc{i}\,D absorption model for all three quiescent bins and the intermediate bin at $2 \leq z < 3$. In the star-forming bin at $3 \leq z < 4$, a model combining Na\,\textsc{i}\,D absorption with a He\,\textsc{i} P-Cygni component is preferred, suggesting the coexistence of ionized and neutral gas outflow phases in this bin. As shown in Figure~\ref{fig:fit}, the selected models reproduce the observed absorption profiles well.

\subsection{Enhanced Neutral Gas Outflows in Quiescent Galaxies}

For the five galaxy bins with significant Na\,\textsc{i}\,D absorption detections, we derive the outflow velocity, neutral gas mass, mass outflow rate, kinetic energy outflow rate, and mass loading factor; the results are summarized in Table~\ref{table:props}. The detected outflows show clear systematic differences across galaxy populations, with quiescent galaxies exhibiting larger values for all derived quantities than non-quiescent detections.

Quiescent galaxies show the highest outflow velocities, reaching $v_{\rm out} = 574^{+77}_{-67}$ and $505^{+54}_{-49}$~km~s$^{-1}$ in the $2 \leq z < 3$ and $3 \leq z < 4$ bins, respectively, approximately twice the velocities measured in the star-forming ($3\leq z<4$, $324^{+103}_{-68}$~km~s$^{-1}$) and intermediate ($2\leq z<3$, $230^{+49}_{-41}$~km~s$^{-1}$) stacks. The Na\,\textsc{i}\,D velocity offsets in all three quiescent bins more than satisfy our criterion for robust outflows ($v_{\rm NaD} \leq -100$~km~s$^{-1}$), confirming that the absorption features are consistently blueshifted. In the $4 \leq z < 5$ bin, the characteristic outflow velocity is somewhat lower, $v_{\rm out} = 313^{+64}_{-53}$~km~s$^{-1}$, due to a narrower velocity dispersion, though a clearly blueshifted absorption profile is still present.

The neutral gas masses and mass outflow rates are likewise highest in the quiescent population. The neutral gas masses in the quiescent bins reach $\log(M_{\rm neutral}/M_\odot) = 7.70^{+0.16}_{-0.14}$,
$7.64^{+0.12}_{-0.09}$, and
$7.43^{+0.13}_{-0.11}$
at $2 \leq z < 3$, $3 \leq z < 4$, and $4 \leq z < 5$, respectively, exceeding the $6.95^{+0.13}_{-0.11}$ and $7.05^{+0.15}_{-0.14}$ measured in the detected star-forming ($3 \leq z < 4$) and intermediate ($2 \leq z < 3$) stacks by $\sim$0.5--0.7~dex. Similarly, the quiescent bins show
$\log(\dot{M}_{\rm out}/M_\odot\,{\rm yr}^{-1}) =
1.47^{+0.15}_{-0.13}$, $1.36^{+0.10}_{-0.09}$, and
$0.94^{+0.16}_{-0.15}$, exceeding the 
$0.46^{+0.20}_{-0.15}$ and $0.42^{+0.17}_{-0.16}$ measured in the detected star-forming ($3 \leq z < 4$) and intermediate ($2 \leq z < 3$) stacks by $\sim0.5$--1 dex. Together, these measurements show that quiescent galaxies host faster and more massive neutral gas outflows.

The kinetic energy outflow rate of the neutral gas outflows follows the same overall trend. Quiescent galaxies exhibit $\log(\dot{E}_{\rm out}/{\rm erg\,s}^{-1}) =
42.5^{+0.2}_{-0.2}$, $42.3^{+0.1}_{-0.1}$, and
$41.4^{+0.3}_{-0.3}$ across the three redshift bins, exceeding the $41.0^{+0.4}_{-0.3}$ and $40.7^{+0.3}_{-0.3}$ measured in the detected star-forming ($3 \leq z < 4$) and intermediate ($2 \leq z < 3$) systems by approximately 0.5--2~dex. The highest kinetic energy outflow rate is measured in the $2\leq z <3$ quiescent bin, reflecting the combination of large neutral gas masses and high outflow velocities. Even at $4\leq z <5$, where the characteristic outflow velocity decreases, the kinetic energy outflow rate remains substantially elevated relative to the non-quiescent detections, indicating that quiescent galaxies host the most energetic neutral gas outflows in our sample.

The same trend is seen in the mass loading factor, $\eta$. While the star-forming ($3 \leq z < 4$) and intermediate ($2 \leq z < 3$) detections yield $\log \eta = -0.69^{+0.70}_{-0.71}$ and $-0.09^{+0.75}_{-0.90}$, respectively, corresponding to $\eta \lesssim 1$, the quiescent bins show $\log \eta = 3.45^{+5.99}_{-2.04}$, $1.68^{+1.64}_{-1.18}$, and $1.45^{+4.21}_{-0.74}$ across the three redshift bins, 2--4~dex higher.  Figure~\ref{fig:eta_comparison} compares our stacked $\eta$ values with individual quiescent and post-starburst systems from \citet{Davies_2024,Valentino_2025, taylor2026jwstexcelssurveyoutflows} and \citet{sun2026censusnadtracedneutral}. Our quiescent $\eta$ values fall within or above the range spanned by independent literature measurements, all of which lie systematically above $\eta=1$. By contrast, the star-forming sample from the literature lies near or below unity, consistent with our own star-forming and intermediate populations. This agreement across independently selected samples indicates that the elevated $\eta$ measured here reflects a genuine population-level property of quiescent galaxies. Given their very low current star formation rates, such large $\eta$ values cannot be readily explained by stellar feedback alone, implying that an additional driving mechanism is required. By contrast, the $\eta$ values measured in the star-forming and intermediate populations are of order unity or below, consistent with neutral gas outflows driven by star formation \citep{Veilleux_2005,Rupke_2005b}.

Taken together, the detected neutral gas outflows in quiescent galaxies are systematically faster, more massive, and more energetic than those detected in the star-forming and intermediate populations, despite their much lower current star formation rates. This trend points to a close connection between powerful neutral gas outflows and the quenching process.

\section{Discussion} \label{sec:discussion}

\subsection{Can Stellar Feedback Explain the Outflows?} \label{sec:discussion_stellar}

\begin{figure}[t]
    \centering
    \includegraphics[width=\columnwidth]{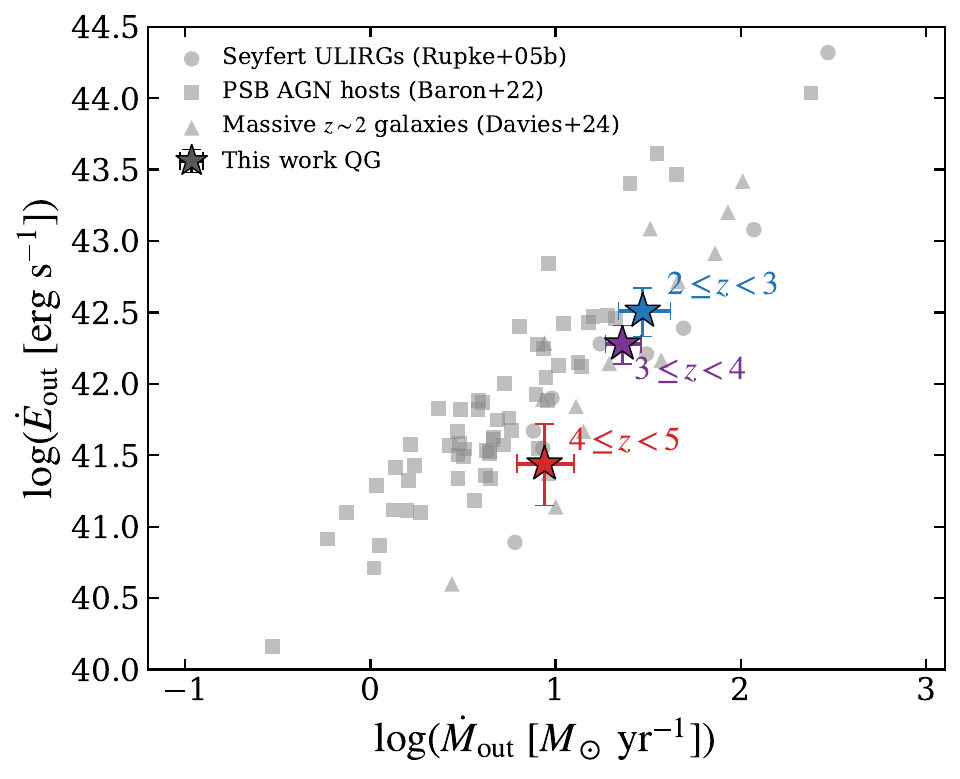}
    \caption{Comparison of the neutral gas mass outflow rate, $\dot{M}{\rm out}$, and kinetic energy outflow rate, $\dot{E}{\rm out}$, for the quiescent bins with AGN-associated literature samples of neutral gas outflows. Stars show the quiescent galaxy stacks in this work, with error bars indicating the 16th--84th percentile ranges. The comparison samples include Seyfert ULIRGs from \citet{Rupke_2005b}, local post-starburst AGN hosts from \citet{Baron_2021}, and massive $z\sim2$ galaxies with Na\,\textsc{i}\,D outflows interpreted as AGN-driven from \citet{Davies_2024}. At a fixed mass outflow rate, the kinetic energy outflow rate depends on the square of the outflow velocity, such that the vertical scatter primarily reflects intrinsic differences in outflow velocity. The quiescent bins lie within the broad region occupied by these literature systems, indicating that their neutral gas outflow mass and energy rates are comparable to those of powerful AGN-associated feedback systems. This comparison alone does not uniquely identify the driving mechanism, but, together with their low current SFRs and the star-formation energy-budget argument, it supports an AGN-related origin.}
    \label{fig:mouteout_scatter}
\end{figure}

A central tension emerges from our results. The galaxies with the least ongoing star formation host the most energetic neutral gas outflows. This raises a fundamental question: how can such powerful outflows be sustained in galaxies with strongly suppressed recent star formation? We therefore examine whether stellar feedback alone can account for the observed outflow properties.

The mass loading factors in the quiescent bins exceed those of the star-forming and intermediate detections by 2--4\,dex, while the current SFRs are extremely low, approaching zero in some redshift bins. Because $\eta$ by definition compares the outflow mass to the star formation presumed to drive it, values this large require either an implausibly efficient coupling of stellar energy to the ISM, or an energy source unrelated to current star formation \citep{Baron_2021,Davies_2024,taylor2026jwstexcelssurveyoutflows,Sun_2026}. In contrast, the star-forming and intermediate detections yield $\eta \lesssim 1$, consistent with the expectations for stellar-feedback-driven winds.

This interpretation is further supported by the recent SFR ratios shown in Figure~\ref{fig:sfr_ratio}. The quiescent bins show negative $\log(\mathrm{SFR_{10}/SFR_{100}})$ medians, indicating that the star formation rate averaged over the past 10 Myr is lower than that averaged over the past 100 Myr, confirming that star formation has been declining on short timescales in these systems. By contrast, the star-forming population shows stable or rising recent star formation. Therefore, the powerful neutral gas outflows in quiescent galaxies cannot be attributed to ongoing star formation activity.

An additional argument is provided by the energy budget shown in Figure~\ref{fig:EoutvsLSF}. We compare the kinetic energy outflow rate inferred from the Na\,\textsc{i}\,D absorption with the mechanical energy injection rate expected from supernova feedback, $\dot{E}_\mathrm{out} = 7\times10^{41}\,\mathrm{SFR}$, based on the Starburst99 calibration adopted by \citet{Veilleux_2005}. While the detected star-forming and intermediate bins are broadly consistent with the stellar-feedback expectation within the systematic uncertainties, all three quiescent bins lie systematically above the relation. We note, however, that the $4 \leq z < 5$ bin carries substantially larger uncertainties than the other two quiescent bins, so it cannot be considered securely above the relation. Nevertheless, all three quiescent bins, including the $4 \leq z < 5$ bin, lie above the relation, and the two lower-redshift bins show a particularly robust offset, indicating that present-day star formation rates are insufficient to account for the observed energetics of neutral gas outflows in these systems, thereby supporting the interpretation based on large mass loading factors.

While the absolute values of $\eta$ depend on assumptions such as the outflow radius, ionization correction, and sodium abundance, the overall trend that quiescent galaxies host outflows with $\eta \gg 1$ is robust. We verify this explicitly by varying the assumed outflow radius over $r_{\rm out}=0.5$--$3$ kpc, which shifts $\eta$ by only $-0.3$ to $+0.5$ dex relative to our fiducial $r_{\rm out}=1$ kpc; all three quiescent bins nevertheless remain above $\eta=1$. The relative ordering across populations likewise remains unchanged regardless of these systematics. Taken together, the observed outflow properties in quiescent galaxies are difficult to reconcile with stellar feedback as the sole driving mechanism.

\subsection{Non-Stellar Origins of the neutral gas outflows} \label{sec:discussion_nonstellar}

If stellar feedback alone cannot account for the observed neutral gas outflows, an additional energy source is required. While the present data do not allow us to directly identify the driving mechanism, the emission-line diagnostics provide useful constraints on the possible origins.

Figure~\ref{fig:bpt_2d_histo} shows the BPT joint distributions for each galaxy population and redshift bin using $\log(\mathrm{[N\,\textsc{ii}]/H\alpha})$ and $\log(\mathrm{[O\,\textsc{iii}]/H\beta})$. Here, the joint distribution for each bin represents all possible combinations of the two line ratios across the individual-galaxy posterior distributions in that bin. Among the quiescent populations, the two statistically reliable bins at $2 \leq z < 3$ and $3 \leq z < 4$ show joint distributions preferentially located above the empirical demarcation of \citet{Kauffmann_2003}, with the majority of individual galaxies lying in the same region. Photoionization by young stars alone typically produces line ratios below this demarcation even in high-sSFR systems; the elevated ratios in the quiescent bins therefore suggest that purely star-forming excitation is unlikely to dominate and are consistent with a harder ionizing source, such as AGN activity or fast shocks. The $4 \leq z < 5$ bin contains too few galaxies with simultaneous line detections to permit statistical conclusions and should be interpreted with caution.

This trend is not driven by a small number of outliers but is reflected in the overall population distributions. Figure~\ref{fig:nii-ha} shows that quiescent galaxies have systematically higher $\log(\mathrm{[N\,\textsc{ii}]/H\alpha})$ distributions than the other populations, and Figure~\ref{fig:oiii-hb} shows elevated $\log(\mathrm{[O\,\textsc{iii}]/H\beta})$ ratios in the quiescent bins at $2 \leq z < 4$, where the joint detections are statistically reliable; the $4 \leq z < 5$ bin is based on far fewer galaxies and carries much larger uncertainties, and is therefore not used to support this trend (see Appendix~\ref{app:bpt} for details). Overall, the shift in the BPT joint diagram reflects a population-wide property of quiescent galaxies rather than the influence of individual extreme objects.

Also, we compare the neutral gas outflow mass and kinetic energy outflow rates of our quiescent bins with literature samples in which AGN activity is either selected or strongly implicated: Seyfert ULIRGs from \citet{Rupke_2005b}, local post-starburst AGN hosts from \citet{Baron_2021}, and massive $z\sim2$ galaxies with Na\,\textsc{i}\,D outflows interpreted as AGN-driven from \citet{Davies_2024} in Figure~\ref{fig:mouteout_scatter}. The quiescent bins occupy the same region of the $\dot{M}_{\rm out}$--$\dot{E}_{\rm out}$ plane as these systems. This comparison alone does not uniquely determine the driving mechanism, especially because some comparison samples may include both AGN and starburst contributions. Nevertheless, combined with the low current SFRs of the quiescent bins and their offset from the star-formation energy scaling established above, the comparable outflow energetics support an AGN-related origin.

The dynamical timescale of the outflow can be estimated as $t_{\rm dyn} = r_{\rm out}/v_{\rm out}$. Adopting the same fiducial $r_{\rm out} = 1$ kpc used above and the characteristic outflow velocities measured in the quiescent bins, we obtain $t_{\rm dyn} \sim 1$--$2$ Myr, shorter than the typical timescales of episodic AGN activity, generally thought to operate on scales of several to tens of Myr \citep{Hopkins_2005}. Because this estimate scales linearly with the assumed outflow radius, the absolute value of $t_{\rm dyn}$ should be interpreted as uncertain. It is therefore possible that, in addition to ongoing AGN feedback, neutral gas accelerated by past AGN activity remains observable at the present epoch \citep{Zubovas_2020} as a relic, or fossil feedback, of AGN activity that has since faded. Consistent with this picture, \citet{Sun_2026} reported a Na\,\textsc{i}\,D outflow in a quiescent galaxy that cannot be accounted for by current star formation or AGN activity alone and interpreted it as a fossil outflow driven by past AGN activity. Ongoing AGN feedback and the remnants of past AGN activity are not necessarily mutually exclusive scenarios; in individual galaxies, both processes may contribute simultaneously to the observed neutral gas outflows. Our stacked data cannot distinguish between these two scenarios on a galaxy-by-galaxy basis, but both point to AGN-related energy injection as the more plausible origin relative to star formation.

Taken together, the BPT distributions and the energetically powerful neutral gas outflows of quiescent galaxies in our sample are consistent with a contribution from ongoing or fossil AGN activity. The present data do not permit direct detection of AGN or definitive identification of the origins of the observed outflows, and contributions from shock excitation cannot be excluded. Diffuse ionized gas could in principle affect BPT line ratios, but such signatures appear to be rare at high redshift and are therefore unlikely to dominate our stacked spectra \citep{hutchison2025jwstwazarc}. Rather than advocating for a specific mechanism, we conclude that the observed outflows are inconsistent with stellar feedback as the sole driver and that their origin is most plausibly connected to non-stellar feedback processes.

\subsection{Implications for Galaxy Quenching} \label{sec:discussion_quenching}

A common picture treats quenching as largely complete once a galaxy has exhausted or expelled its cold star-forming gas, after which it settles into a passive state with little further evolution of its gas content. However, growing evidence from quiescent and post-starburst galaxies at intermediate redshift suggests that substantial gas reservoirs can persist after quenching \citep{Spilker_2018}. The neutral gas masses inferred from the Na\,\textsc{i}\,D absorption (Table~\ref{table:props}) show that significant neutral gas remains in quiescent galaxies across $z = 2$--$5$, and the consistently blueshifted absorption profiles (Section~\ref{sec:results}) demonstrate that this gas is not sitting inertly in the galaxies but is being actively driven outward.

The combination of results from Sections~\ref{sec:discussion_stellar} and~\ref{sec:discussion_nonstellar} suggests that this outflowing gas continues to operate beyond the initial establishment of the quiescent state, and may contribute to sustaining it. Because the mass loading factors in the quiescent bins are far too large to be powered by the residual star formation, and because the emission-line ratios and outflow dynamical timescales point to an additional non-stellar contribution, potentially including ongoing or fossil AGN activity, our results are consistent with a picture in which AGN-driven removal of gas continues after the bulk of star formation has already ceased. Under this interpretation, quenching is not a single event but may instead be followed by an ongoing process of gas regulation, consistent with a maintenance-mode quenching scenario \citep{Beckmann2017,Patil2026}.

This interpretation is qualitatively consistent with cosmological simulations in which AGN feedback contributes to the quenching of massive galaxies and helps maintain quiescence by regulating the subsequent gas supply \citep{Weinberger_2018, Zinger_2020, kurinchivendhan2024originstarformationquenching}. In this picture, feedback can operate through both ejective and preventive channels, linking gas removal from the galaxy to the longer-term suppression of cooling and reaccretion. The neutral gas outflows detected here in quiescent galaxies at $z=2$--5 therefore fit naturally with both a possible AGN contribution and a maintenance-mode scenario in which gas regulation continues after star formation has declined.

Our results also provide a statistical context for a set of recent individual detections of Na\,\textsc{i}\,D outflows in high-redshift galaxies. \citet{taylor2026jwstexcelssurveyoutflows} reported evidence of Na\,\textsc{i}\,D outflows in quiescent galaxies at $z \sim 1.5-5$, but these detections were limited to a handful of individual objects. Our stacking analysis extends this result by showing that such outflows are detected in all three quiescent redshift bins across a sample 3 times larger, indicating they are not restricted to a biased subset of rare or unusual galaxies. \citet{zhu2026againneutraloutflowsz35} detected both neutral inflow and outflow signatures in a single quiescent galaxy at $z \sim 3.5$. Our detection of a similarly declining recent star formation history, alongside energetic outflows, in the $3 \leq z < 4$ quiescent bin (Section \ref{sec:results}) suggests that this combination may be a common, rather than exceptional, feature of quiescent galaxies at this epoch. \citet{Davies_2024} demonstrated that AGN-driven neutral gas outflows are widespread among massive $z \sim 2$ galaxies. We directly compare our quiescent-bin outflow energetics with their samples in Figure \ref{fig:mouteout_scatter} and find that the two occupy a similar region of the $\dot{M}_{\rm out}$--$\dot{E}_{\rm out}$ plane. The present work extends these findings from individual systems to a statistical sample. Recently, \citet{sapori2026outflowsearlyuniverseneutral} used a large stacking analysis of low-resolution prism spectroscopy to show that Na\,\textsc{i}\,D detections in quiescent galaxies are statistically common. However, the limited resolution of the prism prevented direct measurement of blueshifts in individual absorption profiles, so their outflow interpretation relied on an indirect extrapolation from literature equivalent-width (EW)–velocity relations. Using medium-resolution spectroscopy, our work directly measures outflow velocities, masses, and kinetic energies for each stack without this limitation, adding direct kinematic grounding to the trend they establish statistically. 

While our stacking approach cannot isolate the outflow-driving mechanism in any individual galaxy, the statistical prevalence of these outflows and their extreme mass loading factors provides a new observational constraint that quenching models must accommodate. In quiescent galaxies at these redshifts, the suppression of star formation appears to be accompanied by continued, energetically significant redistribution or removal of neutral gas well after star formation has already declined.

\subsection{Robustness of the Main Results}
\label{sec:discussion_robustness}

The results presented in Sections~\ref{sec:discussion_stellar}--\ref{sec:discussion_quenching} rely on the adopted galaxy classification and on a heterogeneous JWST sample in which stellar mass and spectral quality are not fully independent. We therefore performed additional robustness tests to assess whether our main conclusions depend on the galaxy-classification scheme, or other potential drivers such as stellar mass.

The results in Sections~\ref{sec:discussion_stellar}--\ref{sec:discussion_quenching} use a classification of galaxies based on the sSFR averaged over the most recent 10 Myr. Because this window is short relative to the full star formation history of a galaxy, we further test the classification using an independent indicator, $t_{90}$, the lookback time at which 90\% of the stellar mass had formed, which reflects the long-term formation history rather than the recent sSFR (Appendix~\ref{app:t90_robustness}). A substantial fraction of the galaxies originally classified as intermediate based on sSFR are found to be quiescent-like under this indicator, which aligns naturally with the weaker and less coherent Na\,\textsc{i}\,D signals observed in the intermediate bins relative to the quiescent bins.

After reconstructing the sample using the $t_{90}$-based classes, Na\,\textsc{i}\,D absorption remains most robustly detected in the quiescent bins, and the derived outflow velocities and mass loading factors are consistent with the original sSFR-based values. The intermediate population, however, no longer yields a significant detection in any redshift bin, precluding a direct comparison of the outflow hierarchy across all three populations. Nevertheless, the extreme mass loading factors measured in the quiescent bins are robust to this change in classification method.


Because the present JWST sample is heterogeneous, we also examined whether stellar mass could account for the observed trends. Stellar mass and spectral quality are not fully independent, and the quiescent population is concentrated among higher-mass, higher-S/N systems (Appendix~\ref{app:mass_robustness}). First, excluding galaxies with $M_\star < 10^9\,M_\odot$ leaves the detected outflow properties essentially unchanged, since this cut removes only a small number of star-forming galaxies and no intermediate or quiescent systems. Second, splitting the sample at the median stellar mass, $\log M_\star = 9.99$, shows that Na\,\textsc{i}\,D outflows are detected almost exclusively in the high-mass half, consistent with its systematically higher continuum S/N; no quiescent stack could be constructed at $z \ge 3$ in the low-mass half. Within the high-mass half, however, quiescent galaxies retain mass loading factors 2--4 dex above the intermediate and star-forming detections at comparable mass and redshift. We therefore conclude that stellar mass affects the detectability of the Na\,\textsc{i}\,D signal, but cannot by itself explain the elevated mass loading factors observed in quiescent galaxies.

Taken together, these tests demonstrate that the principal conclusions of this work, particularly the unusually large mass loading factors in quiescent galaxies and their inconsistency with stellar feedback as the sole driving mechanism, are insensitive to the adopted galaxy-classification scheme and cannot be explained solely by stellar-mass-dependent selection effects.

\subsection{Limitations and Future Work}

As this work is based on spectral stacking, the diversity of individual galaxies cannot be assessed directly. The derived mass outflow rates and mass loading factors are also subject to systematic uncertainties arising from assumptions about the outflow radius, geometry, ionization correction, and sodium abundance. However, since identical assumptions are applied to all stacked bins, these uncertainties are unlikely to affect the relative comparisons across galaxy populations.

Also, stellar mass and spectral quality are not fully independent in the present sample, and non-detections in the low-mass and high-redshift bins should be interpreted cautiously. This is further complicated by the heterogeneity of the parent catalog, which aggregates spectra from many independent JWST programs with different target selection strategies and observing depths, a heterogeneity that our mass and S/N tests do not fully capture.

We also caution that neither the sSFR-based nor the $t_{90}$-based
classification distinguishes between recently quenched, post-starburst systems and long-quenched, passively evolving galaxies within the quiescent bin. This is also reflected in the broad distribution of recent-to-past star formation ratios in Figure~\ref{fig:sfr_ratio}, where the quiescent bins have negative median values but include a tail toward higher $\mathrm{SFR}_{10}/\mathrm{SFR}_{100}$. By definition, post-starburst galaxies are quiescent systems that quenched only recently, and can retain long-term star formation history indicators consistent with an earlier, extended star-forming phase. It is therefore likely that a sub-population within our quiescent bin consists of such recently quenched systems rather than galaxies that have been passive for a long time. Because post-starburst galaxies are known to host some of the most extreme neutral gas outflows observed at both low and high redshift \citep{Tremonti_2007,Baron_2021,Belli_2024,taylor2026jwstexcelssurveyoutflows}, this sub-population composition would tend to reinforce rather than bias against the extreme mass loading factors we measure in the quiescent bins. Disentangling the relative contributions of recently quenched and long-quenched galaxies within the quiescent population would require additional diagnostics.

Larger JWST spectroscopic samples and spatially resolved spectroscopy with IFU-like observations will be essential for directly constraining the outflow structure and driving mechanism in individual quiescent galaxies. Furthermore, by resolving the spatial distribution and kinematics of neutral gas outflows in individual systems, such observations will provide deeper insight into their physical origin and evolution. Complementary constraints on the cold gas reservoirs of these galaxies could come from ALMA observations of molecular gas \citep{Bezanson_2022}, which trace a different phase of the gas cycle than the neutral atomic gas probed by Na\,\textsc{i}\,D, and would help establish whether the extreme outflows reported here are depleting or being replenished by molecular gas. Strongly lensed galaxies may provide a complementary route toward resolving individual outflows, since gravitational magnification can enable clump-scale measurements of neutral gas inflows and outflows (R. Oh et al., in preparation); such analyses are currently underway as part of the JWST LEGGOS survey \citep{Khullar_2021,klein2024coollampsvilensmodel,hutchison2025jwstwazarc,khullar2026leggosijwstleggos, abedi2026leggosiistronglens,ross2026leggosiiimappingstar}. Combined with independent AGN diagnostics at X-ray, radio, and mid-infrared wavelengths, they will help clarify the causal relationship between neutral gas outflows and AGN activity.

Together, these future efforts will provide more quantitative
constraints on the origin of neutral gas feedback and its role in the
quenching of galaxies at high redshift.

\section{Conclusion} \label{sec:conclusion}


We perform an absorption stacking analysis on 274 galaxies at $z=2$--5, spanning star-forming, intermediate, and quiescent populations. The sample combines medium-resolution JWST/NIRSpec G140M, G235M, and G395M spectroscopy from the DAWN JWST Archive and the \textsc{ember} JWST survey program with NIRCam photometry.

After removing the stellar contribution to the Na\,\textsc{i}\,D feature, we present the statistical detection of neutral gas outflows traced by the residual, purely interstellar Na\,\textsc{i}\,D absorption.

Na\,\textsc{i}\,D absorption is detected in all three quiescent galaxy bins, while only isolated detections are found in the star-forming ($3 \leq z < 4$) and intermediate ($2 \leq z < 3$) populations. Quiescent galaxies show the highest outflow velocities, neutral gas masses, mass outflow rates, kinetic energy outflow rates, and mass loading factors of all populations.

The combination of extreme mass loading factors and declining recent star-formation histories in the quiescent bins indicates that the observed outflows are difficult to explain by the level of current residual star formation alone. The BPT distributions in the statistically reliable quiescent bins at $2 \leq z < 3$ and $3 \leq z < 4$ further motivate consideration of multiple non-stellar ionization sources, including AGN photoionization and fast shocks. The short dynamical timescales of the outflows relative to typical AGN duty cycles remain consistent with either ongoing or fossil AGN activity, although the present stacked spectra do not uniquely identify the dominant driving mechanism.

These results suggest that high-redshift quiescent galaxies are not passive, gas-exhausted systems, but rather host substantial reservoirs of outflowing neutral gas well after star formation has largely ceased. This provides new observational constraints on the quenching of galaxies in the early Universe, and motivates future large-scale JWST spectroscopic campaigns to clarify the physical origin of these outflows and the relative roles of AGN activity and shocks.

\section{Acknowledgments}

RO would first like to express her sincere gratitude to her advisors, Arianna S. Long and Gourav Khullar, for their invaluable guidance, support, and encouragement throughout this research. She is also deeply grateful to her family, including Bangsil Kim, Junhyeong Oh, Okhyang Cho, Haengik Oh, Yeongyeong Oh, Mirae Oh, and Yull Oh, for their unwavering support. She also thanks her three dear friends, Haeli Kim, Youngsil Han, and Jina Lee, for always encouraging her. She also thanks Professor Kim (KAIST) for his advice and encouragement in pursuing the research opportunity at the University of Washington. Finally, she extends her gratitude to everyone who offered help and inspiration throughout this research, especially the many wonderful people she met at the University of Washington.

GK notes that we do not use the full name of JWST due to the person after whom this telescope is named and their role as NASA administrator during the ``Lavender Scare'', as per the $\#RenameJWST$ protest movement. GK would like to thank the Baum Grant and Fellowship at the University of Washington for support during this work, as well as the ALMA Ambassador Program (administered by NAASC and NRAO). GK would also like to thank the DiRAC Institute in the Department of Astronomy at the University of Washington. The DiRAC Institute is supported through generous gifts from the
Charles and Lisa Simonyi Fund for Arts and Sciences, Janet and Lloyd Frink, and the Washington Research Foundation. 

This work is based on observations made with the NASA/ESA/CSA James Webb Space Telescope, obtained at the Space Telescope Science Institute, which is operated by the Association of Universities for Research in Astronomy, Incorporated, under NASA contract NAS5-03127. Some of the data products presented herein were retrieved from the Dawn JWST Archive (DJA). DJA is an initiative of the Cosmic Dawn Center (DAWN), which is funded by the Danish National Research Foundation under grant DNRF140.
We acknowledge the original JWST programs contributing to the DJA data used in this work: Programs 1180, 1181, 1207, 1210, 1211, 1212, 1286, 1287, 1345, 1671, 1810, 1879, 1914, 3215, 3543, 3567, 4233, and 4446. This work also makes use of data from the \textsc{ember} JWST survey; support for program number GO-7076 was provided through a grant from the STScI under NASA contract NAS5-03127. 

This work used Anvil at Purdue University through allocation \#PHY260106 from the Advanced Cyberinfrastructure Coordination Ecosystem: Services \& Support (ACCESS) program \citep{10.1145/3569951.3597559}, which is supported by U.S. National Science Foundation grants \#2138259, \#2138286, \#2138307, \#2137603, and \#2138296. Anvil is supported by the National Science Foundation under Grant No. 2005632 \citep{10.1145/3491418.3530766}.

This research made use of Python and the following open-source software packages:
Astropy \citep{2013A&A...558A..33A, 2018AJ....156..123A, 2022ApJ...935..167A}, NumPy \citep{Harris_2020}, SciPy \citep{Virtanen_2020}, pandas \citep{mckinney-proc-scipy-2010,reback2020pandas}, Matplotlib \citep{4160265}, Prospector \citep{Johnson_2021}, FSPS \citep{2010ascl.soft10043C}, emcee \citep{Foreman_Mackey_2013}, corner.py \citep{2016JOSS....1...24F}, and spectres \citep{carnall2017spectresfastspectralresampling}.

\bibliography{Ref}{}
\bibliographystyle{aasjournalv7}

\appendix

\section{One-Dimensional Line-Ratio Distributions Underlying the BPT Diagram} \label{app:bpt}

The joint distribution of $\log([\mathrm{N II}]/\mathrm{H}\alpha)$ and $\log([\mathrm{O III}]/\mathrm{H}\beta)$ shown in Figure~\ref{fig:bpt_2d_histo} is constructed from the one-dimensional posterior distributions of each line ratio for individual galaxies. Here we present these two line ratios separately as pooled one-dimensional distributions for each galaxy population and redshift bin, following the same Monte Carlo procedure used in Figure~\ref{fig:sfr_ratio}. For each bin, we combine Monte Carlo realizations that propagate the individual posterior uncertainties on the line ratio and report the pooled median together with the median $1\sigma$ uncertainty.

\begin{figure}
    \centering
    \includegraphics[width=\textwidth]{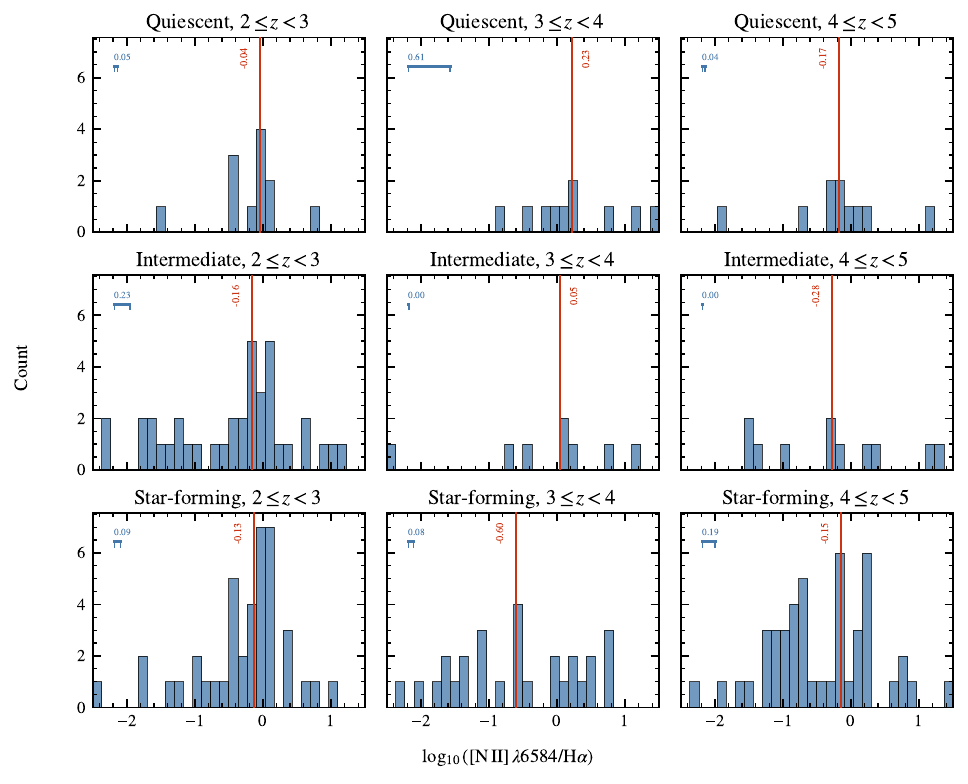}
    \caption{Distribution of $\log([\mathrm{N II}]\,\lambda6584/\mathrm{H}\alpha)$ for each galaxy population and redshift bin, corresponding to the horizontal axis of the BPT diagram in Figure~\ref{fig:bpt_2d_histo}. In each panel, the histogram is constructed from Monte Carlo realizations that propagate the individual posterior uncertainties on $\log([\mathrm{N II}]/\mathrm{H}\alpha)$. The red vertical line marks the pooled median, with its value annotated in red; the blue horizontal bar indicates the median $1\sigma$ uncertainty on the ratio. Quiescent galaxies show systematically higher $\log([\mathrm{N II}]/\mathrm{H}\alpha)$ values than the star-forming population in the statistically reliable bins, consistent with the offset seen in the joint BPT distribution of Figure~\ref{fig:bpt_2d_histo}.}
    \label{fig:nii-ha}
\end{figure}

Figure~\ref{fig:nii-ha} shows the resulting distributions of $\log([\mathrm{N II}]/\mathrm{H}\alpha)$. In the bins with the most reliable joint detections in Figure~\ref{fig:bpt_2d_histo}, the quiescent population shows pooled medians of $-0.04$ and $0.23$ at $2 \leq z < 3$ and $3 \leq z < 4$, respectively, systematically higher than the star-forming medians of $-0.13$ and $-0.60$ in the same redshift bins. The intermediate population at $2 \leq z < 3$ has a median of $-0.16$, consistent with its classification between the star-forming and quiescent populations. In these bins, the median $1\sigma$ uncertainty is generally below $0.6$, so the trend toward higher $\log([\mathrm{N II}]/\mathrm{H}\alpha)$ in quiescent galaxies is comparatively robust.

\begin{figure}[h]
    \centering
    \includegraphics[width=\textwidth]{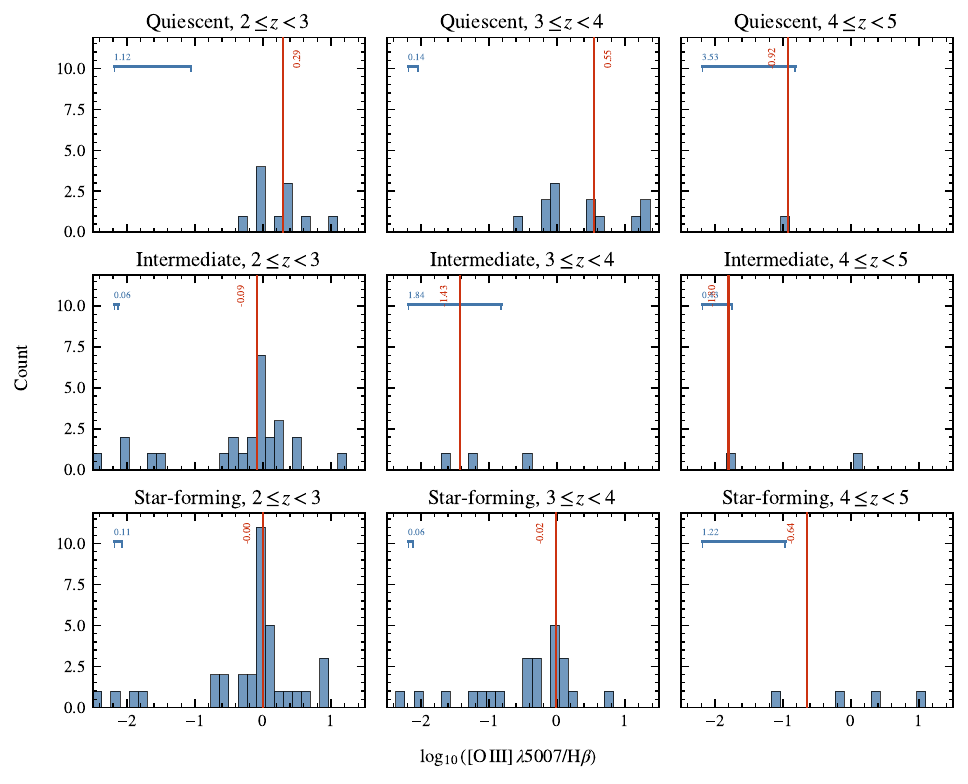}
    \caption{Distribution of $\log([\mathrm{O III}]\,\lambda5007/\mathrm{H}\beta)$ for each galaxy population and redshift bin, corresponding to the vertical axis of the BPT diagram in Figure~\ref{fig:bpt_2d_histo}. As in Figure~\ref{fig:nii-ha}, the histogram in each panel is constructed from Monte Carlo realizations that propagate the individual posterior uncertainties on $\log([\mathrm{O III}]/\mathrm{H}\beta)$, with the red vertical line marking the pooled median and the blue horizontal bar indicating the median $1\sigma$ uncertainty. Quiescent galaxies show elevated $\log([\mathrm{O III}]/\mathrm{H}\beta)$ ratios relative to the star-forming population in the statistically reliable bins. However, the uncertainties are generally larger than for $\log([\mathrm{N II}]/\mathrm{H}\alpha)$, particularly in bins with few jointly detected galaxies. The $4 \leq z < 5$ bins are based on fewer galaxies and should be interpreted with caution.}
    \label{fig:oiii-hb}
\end{figure}

Figure~\ref{fig:oiii-hb} shows the corresponding distributions of $\log([\mathrm{O III}]/\mathrm{H}\beta)$. In the two statistically reliable quiescent bins, the pooled medians ($0.29$ and $0.55$ at $2 \leq z < 3$ and $3 \leq z < 4$, respectively) are higher than the star-forming medians ($-0.00$ and $-0.02$ in the same bins). However, the median uncertainty on $\log([\mathrm{O III}]/\mathrm{H}\beta)$ is generally larger than that on $\log([\mathrm{N II}]/\mathrm{H}\alpha)$, and in bins with fewer jointly detected galaxies ($4 \leq z < 5$) the median $1\sigma$ uncertainty exceeds $3$, so trends in these bins should be interpreted with caution. Bins with fewer than five galaxies with jointly detected line ratios (marked with pink dashed borders in Figure~\ref{fig:bpt_2d_histo}) show particularly large uncertainties and are not used to support the statistical trends discussed in the main text.

Together, these one-dimensional distributions show that the offset of the quiescent population toward the upper-right region of the BPT diagram in Figure~\ref{fig:bpt_2d_histo} is not driven by an imbalance in only one of the two line ratios, but reflects a trend present in both $[\mathrm{N II}]/\mathrm{H}\alpha$ and $[\mathrm{O III}]/\mathrm{H}\beta$ in the statistically reliable bins. Given the larger uncertainty on $\log([\mathrm{O III}]/\mathrm{H}\beta)$, this trend is more firmly supported by $\log([\mathrm{N II}]/\mathrm{H}\alpha)$, with $\log([\mathrm{O III}]/\mathrm{H}\beta)$ providing supporting rather than independent evidence.

\section{Robustness Check with SFH-based Reclassification}
\label{app:t90_robustness}

Here, we examine the internal structure of the sSFR-based intermediate classification using an independent, SFH-based indicator, $t_{90}$. The quantity $t_{90}$ is defined as the lookback time at which 90\% of the stellar mass had formed, derived from the non-parametric star formation history in the posterior for each galaxy, and reflects the long-term formation history of a galaxy, in contrast to the sSFR averaged over the most recent 10\,Myr used in the primary classification.

We apply $k$-means clustering ($k=3$) to the $t_{90}$ distribution of the full sample, yielding class boundaries of $t_{90} = 0.039$ and $0.198$\,Gyr that separate SFG-like ($t_{90} < 0.039$\,Gyr), INT-like ($0.039 \leq t_{90} < 0.198$\,Gyr), and QG-like ($t_{90} \geq 0.198$\,Gyr) galaxies. We then apply this classification to the galaxies originally identified as intermediate based on sSFR. Of the 74 original intermediate galaxies, 32 (43.2\%) are reclassified as QG-like, 25 (33.8\%) remain INT-like, and only 17 (23.0\%) are classified as SFG-like.

To test whether this reclassification affects the main results, we recomputed the outflow properties after reconstructing the sample based on the $t_{90}$-based classes. The resulting sample sizes per redshift bin and galaxy type are summarized in Table~\ref{table:t90_stack}. Under this reclassification, the intermediate population is reduced to only 2 and 5 galaxies at $3 \leq z < 4$ and $4 \leq z < 5$, respectively, too few to construct a statistically meaningful stack; we therefore do not attempt to interpret the intermediate bins under this classification. By contrast, the quiescent and star-forming bins retain sample sizes that are broadly comparable to or larger than those in the original classification, allowing us to test the robustness of the main results for these two populations.

As shown in Table~\ref{table:t90_vout_eta}, the mass loading factor and outflow velocity in the quiescent bins change only modestly after reclassification, and $\eta > 1$ is maintained in all three redshift bins. The full set of derived physical and outflow properties under the $t_{90}$-based classification is given in Table~\ref{table:t90_full}. Under this reclassification, Na\,\textsc{i}\,D absorption remains most robustly detected in the quiescent bins, with $v_{\rm out}$ and $\eta$ comparable to the original sSFR-based values. Because the intermediate population can no longer be meaningfully tested and the star-forming detection remains limited to $3 \leq z < 4$, a direct three-way comparison of the outflow hierarchy is not possible under this classification. Nevertheless, the central result of this work, that quiescent galaxies host outflows with $\eta \gg 1$ inconsistent with stellar feedback, is unaffected: this result depends only on the quiescent bins, whose sample sizes and derived properties remain robust under the independent $t_{90}$-based classification.

\section{Robustness Check With Stellar Mass}
\label{app:mass_robustness}

Because the present JWST sample is heterogeneous, stellar mass and spectral quality are not fully independent, and the quiescent population identified in Section~\ref{sec:sed} is concentrated among higher-mass systems. Here we examine whether the elevated outflow properties measured in the quiescent bins in Section~\ref{sec:results} could instead be driven by this mass dependence, using two complementary tests.

We first repeat the full stacking and fitting procedure after excluding all galaxies with $M_\star < 10^9 M_\odot$. This cut removes a small number of star-forming galaxies in each redshift bin, but no intermediate or quiescent galaxies, since these populations in our sample are already dominated by higher-mass systems above this threshold. Table~\ref {table:mass_test} lists the resulting sample sizes and derived outflow properties. The outflow velocities and mass loading factors in the detected bins are essentially unchanged from the full-sample values reported in Table~\ref{table:props}. For example, the star-forming detection at $3 \le z < 4$ yields $\log\eta = -0.71$ after this cut, compared to $-0.69$ in the full sample. The quiescent-bin properties remain unchanged by construction, since this cut removed no quiescent galaxies.

\begin{figure*}[t]
    \centering
    \includegraphics[width=\textwidth]{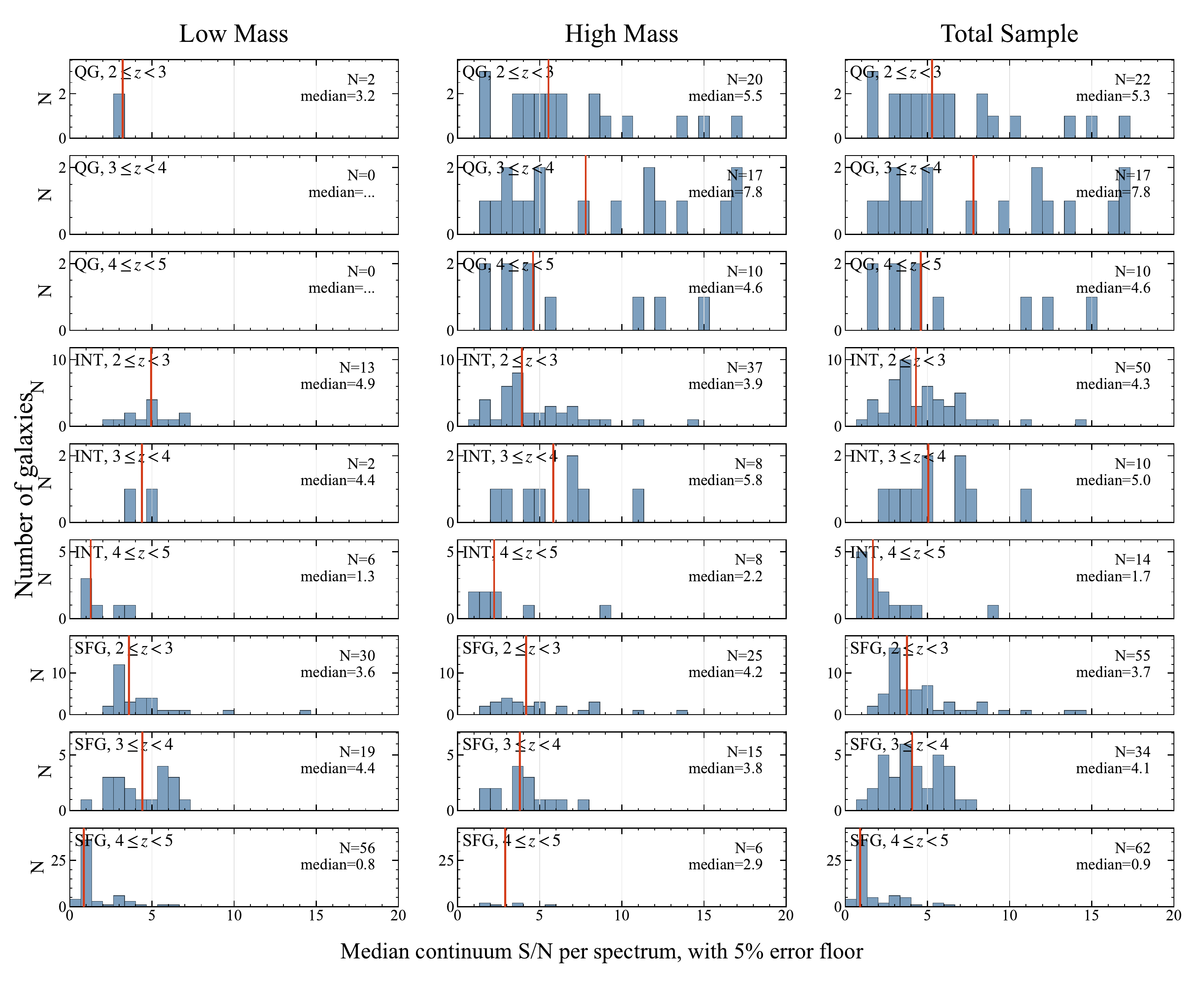}
    \caption{Median continuum S/N per spectrum, with a 5\% error floor applied, for each galaxy type and redshift bin. Panels are arranged in three columns corresponding to the low-mass subsample, the high-mass subsample, and the total sample, with rows separating quiescent, intermediate, and star-forming galaxies at $2 \le z < 3$, $3 \le z < 4$, and $4 \le z < 5$. The red vertical line in each panel marks the median S/N. The high-mass subsample shows systematically higher S/N than the low-mass subsample across nearly all galaxy types and redshift bins, consistent with the concentration of Na\,\textsc{i}\,D detections in the high-mass half.}
    \label{fig:mass_snr}
\end{figure*}

We next split the full sample at its median stellar mass, $\log(M_\star/M_\odot) = 9.99$, into high-mass and low-mass halves and repeat the stacking separately for each, as shown in Tables~\ref{table:mass_high_half} and~\ref{table:mass_low_half}. Na\,\textsc{i}\,D outflows are detected almost exclusively in the high-mass half. No quiescent stack could be constructed at $3 \le z < 5$ in the low-mass half, and the low-mass quiescent bin at $2 \le z < 3$ contains only two galaxies, too few to interpret. This asymmetry is broadly consistent with the continuum signal-to-noise ratio of the individual spectra entering each stack, shown in Figure~\ref{fig:mass_snr}. This indicates that stellar mass strongly affects the detectability of the Na\,\textsc{i}\,D signal in our sample, largely through its association with spectral S/N, rather than reflecting a physical requirement that outflows occur only in massive galaxies.

Critically, within the high-mass half alone, the population-level trend identified in Section~\ref{sec:results} persists. The quiescent bins retain $\log\eta = 3.33$, $1.64$, and $1.62$ across the three redshift bins. These values are systematically higher than the intermediate detection at $2 \le z < 3$, which yields $\log\eta = -0.15$, and the star-forming detections at $2 \le z < 3$ and $3 \le z < 4$, which yield $\log\eta = -0.74$ and $-0.83$, respectively, at comparable stellar mass. Because these comparisons are made within the same high-mass range, the elevated mass loading factors in quiescent galaxies cannot be attributed to a systematic mass difference between the populations.

We conclude that stellar mass and continuum S/N contribute to the detectability of the Na\,\textsc{i}\,D outflow signal in our sample. However, within the available high-mass subsample, the enhanced outflow signatures observed in quiescent galaxies are not solely a mass effect.

\section{Absorption Line Fitting Priors}
\label{app:naid_priors}

Table~\ref{tab:naid_priors} summarizes the free parameters and prior ranges adopted in the absorption line fitting (Section~\ref{sec:fitting}). All priors are uniform within the listed bounds. The MCMC walkers are initialized in a compact distribution around the least-squares best-fit solution.

\FloatBarrier

\FloatBarrier

\begin{table*}[t]
\vspace{4.0em}
\centering

\begin{minipage}{0.92\textwidth}
\centering

\begin{minipage}[t]{0.45\linewidth}
\centering
\caption{Number of Stacked Targets under the $t_{90}$-based Classification}
\label{table:t90_stack}

\begin{tabular*}{\linewidth}{@{\extracolsep{\fill}}lccc}
\hline\hline
Type & $z=2$--3 & $z=3$--4 & $z=4$--5 \\
\hline
Star-forming & 67 & 37 & 64 \\
Intermediate & 18 & 2 & 5 \\
Quiescent & 42 & 22 & 17 \\
\hline
\end{tabular*}

\vspace{0.4em}
\parbox{\linewidth}{\scriptsize
\hangindent=0.8em
\hangafter=1
\noindent \textsc{Note}---Columns give the number of galaxies in each redshift bin
($z=2$--3, $3$--4, $4$--5) classified by spectral type (star-forming,
intermediate, quiescent). Sample sizes correspond to the reclassification
based on $t_{90}$ using $k$-means clustering ($k=3$;
Appendix~\ref{app:t90_robustness}), with class boundaries at
$t_{90}=0.039$ and $0.198\,$Gyr.
}
\end{minipage}%
\hfill%
\begin{minipage}[t]{0.47\linewidth}
\centering
\caption{Comparison of Key Outflow Properties in the Quiescent Bins Before and After $t_{90}$-based Reclassification}
\label{table:t90_vout_eta}

{\setlength{\tabcolsep}{2.0pt}
\begin{tabular*}{\linewidth}{@{\extracolsep{\fill}}ccccc}
\hline\hline
$z$ &
\multicolumn{2}{c}{$v_{\rm out}$ [km s$^{-1}$]} &
\multicolumn{2}{c}{$\log \eta$} \\
 & original & $t_{90}$-based & original & $t_{90}$-based \\
\hline
2--3 & $574^{+77}_{-67}$ & $478^{+69}_{-67}$ & $3.45^{+5.99}_{-2.04}$ & $2.27^{+0.55}_{-0.69}$ \\
3--4 & $505^{+54}_{-49}$ & $495^{+60}_{-52}$ & $1.68^{+1.64}_{-1.18}$ & $1.76^{+0.10}_{-0.87}$ \\
4--5 & $313^{+64}_{-53}$ & $408^{+79}_{-71}$ & $1.45^{+4.21}_{-0.74}$ & $1.25^{+1.09}_{-0.34}$ \\
\hline
\end{tabular*}
}

\vspace{0.4em}
\parbox{\linewidth}{\scriptsize
\hangindent=0.8em
\hangafter=1
\noindent \textsc{Note}---Columns compare the outflow velocity ($v_{\rm out}$) and
mass loading factor ($\log \eta$) in the quiescent bins across redshift,
under the original sSFR-based and $t_{90}$-based classifications. Quoted uncertainties
indicate the 16th and 84th percentiles of the posterior distributions.
Original values are from Table~\ref{table:props}, and $t_{90}$-based values
are from Table~\ref{table:t90_full}.
}
\end{minipage}

\end{minipage}

\vspace{7.0em}
\end{table*}

\begin{deluxetable}{lllccccccccc}
\tabletypesize{\scriptsize}
\tablewidth{0pt}
\tablecaption{Derived Physical and Outflow Properties under the $t_{90}$-based Classification}
\tablehead{
\colhead{Type} &
\colhead{$z$} &
\colhead{$N$} &
\colhead{$\log M_\star$} &
\colhead{$\log {\rm SFR}$} &
\colhead{$|v_{\rm Na\,\textsc{i}\,D}|$} &
\colhead{$\sigma_{\rm Na\,\textsc{i}\,D}$} &
\colhead{$v_{\rm out}$} &
\colhead{$\log M_{\rm neutral}$} &
\colhead{$\log \dot{M}$} &
\colhead{$\log \dot{E}_{\rm out}$} &
\colhead{$\log \eta$} \\
\colhead{} &
\colhead{} &
\colhead{} &
\colhead{[$M_\odot$]} &
\colhead{[$M_\odot\,{\rm yr}^{-1}$]} &
\colhead{[km s$^{-1}$]} &
\colhead{[km s$^{-1}$]} &
\colhead{[km s$^{-1}$]} &
\colhead{[$M_\odot$]} &
\colhead{[$M_\odot\,{\rm yr}^{-1}$]} &
\colhead{[erg s$^{-1}$]} &
\colhead{}
}
\startdata
SFG & 2--3 & 67 & $9.92^{+0.57}_{-0.50}$ & $0.95^{+0.39}_{-0.55}$ & \nodata & \nodata & \nodata & \nodata & \nodata & \nodata & \nodata \\[4pt]
SFG & 3--4 & 37 & $9.95^{+0.57}_{-0.45}$ & $1.16^{+0.56}_{-0.53}$ & $166^{+97}_{-51}$ & $81^{+36}_{-23}$ & $331^{+156}_{-73}$ & $6.95^{+0.14}_{-0.14}$ & $0.46^{+0.27}_{-0.17}$ & $41.0^{+0.6}_{-0.4}$ & $0.11^{+1.69}_{-2.17}$ \\[4pt]
SFG & 4--5 & 64 & $9.23^{+0.63}_{-0.49}$ & $0.80^{+0.46}_{-0.47}$ & \nodata & \nodata & \nodata & \nodata & \nodata & \nodata & \nodata \\[4pt]
INT & 2--3 & 18 & $10.17^{+0.36}_{-0.32}$ & $-0.07^{+0.68}_{-0.89}$ & \nodata & \nodata & \nodata & \nodata & \nodata & \nodata & \nodata \\[4pt]
INT & 3--4 & 2 & $10.46^{+0.19}_{-0.84}$ & $-0.02^{+0.80}_{-0.47}$ & \nodata & \nodata & \nodata & \nodata & \nodata & \nodata & \nodata \\[4pt]
INT & 4--5 & 5 & $9.91^{+0.07}_{-0.29}$ & $-0.22^{+0.73}_{-0.51}$ & \nodata & \nodata & \nodata & \nodata & \nodata & \nodata & \nodata \\[4pt]
QG & 2--3 & 42 & $10.75^{+0.62}_{-0.52}$ & $-1.01^{+0.55}_{-0.42}$ & $95^{+26}_{-21}$ & $189^{+36}_{-34}$ & $478^{+69}_{-67}$ & $7.56^{+0.14}_{-0.12}$ & $1.26^{+0.13}_{-0.14}$ & $42.1^{+0.2}_{-0.2}$ & $2.27^{+0.55}_{-0.69}$ \\[4pt]
QG & 3--4 & 22 & $10.89^{+0.42}_{-0.71}$ & $-0.42^{+0.77}_{-0.00}$ & $104^{+26}_{-22}$ & $196^{+31}_{-29}$ & $495^{+60}_{-52}$ & $7.62^{+0.12}_{-0.09}$ & $1.34^{+0.10}_{-0.10}$ & $42.2^{+0.1}_{-0.2}$ & $1.76^{+0.10}_{-0.87}$ \\[4pt]
QG & 4--5 & 17 & $10.40^{+0.55}_{-0.24}$ & $-0.40^{+0.18}_{-0.91}$ & $222^{+52}_{-56}$ & $92^{+35}_{-26}$ & $408^{+79}_{-71}$ & $7.24^{+0.13}_{-0.14}$ & $0.85^{+0.18}_{-0.16}$ & $41.6^{+0.3}_{-0.3}$ & $1.25^{+1.09}_{-0.34}$ \\[4pt]
\enddata
\tablecomments{Columns are as follows: galaxy type (SFG = star-forming, INT = intermediate, QG = quiescent), redshift bin, number of stacked galaxies ($N$), stellar mass, star formation rate, Na\,\textsc{i}\,D absorption velocity centroid, Na\,\textsc{i}\,D velocity dispersion, outflow velocity, neutral gas mass, mass outflow rate, kinetic energy outflow rate, and mass loading factor ($\eta \equiv \dot{M}_{\rm out}/{\rm SFR}$). Outflow properties are derived from the Na\,\textsc{i}\,D absorption-line stacking and fitting for each galaxy population and redshift bin under the $t_{90}$-based reclassification (Appendix~\ref{app:t90_robustness}). The outflow velocity is defined as $v_{\rm out}=|v_{\rm NaD}|+2\sigma_{\rm NaD}$. Larger values of $|v_{\rm NaD}|$ indicate stronger (more blueshifted) 
outflows. Quoted uncertainties indicate the 16th, 50th, and 84th percentiles of the posterior distributions. Entries marked as \nodata\ correspond to Na\,\textsc{i}\,D non-detections, for which the outflow properties could not be robustly constrained.}
\label{table:t90_full}
\end{deluxetable}

\begin{deluxetable}{lllccccccccc}
\tabletypesize{\scriptsize}
\tablewidth{0pt}
\tablecaption{Derived Physical and Outflow Properties after Removing Galaxies with $M_\star < 10^9\,M_\odot$}
\tablehead{
\colhead{Type} &
\colhead{$z$} &
\colhead{$N$} &
\colhead{$\log M_\star$} &
\colhead{$\log {\rm SFR}$} &
\colhead{$|v_{\rm Na\,\textsc{i}\,D}|$} &
\colhead{$\sigma_{\rm Na\,\textsc{i}\,D}$} &
\colhead{$v_{\rm out}$} &
\colhead{$\log M_{\rm neutral}$} &
\colhead{$\log \dot{M}$} &
\colhead{$\log \dot{E}_{\rm out}$} &
\colhead{$\log \eta$} \\
\colhead{} &
\colhead{} &
\colhead{} &
\colhead{[$M_\odot$]} &
\colhead{[$M_\odot\,{\rm yr}^{-1}$]} &
\colhead{[km s$^{-1}$]} &
\colhead{[km s$^{-1}$]} &
\colhead{[km s$^{-1}$]} &
\colhead{[$M_\odot$]} &
\colhead{[$M_\odot\,{\rm yr}^{-1}$]} &
\colhead{[erg s$^{-1}$]} &
\colhead{}
}
\startdata
SFG & 2--3 & 53 & $9.94^{+0.60}_{-0.41}$ & $0.98^{+0.46}_{-0.34}$ & \nodata & \nodata & \nodata & \nodata & \nodata & \nodata & \nodata \\[4pt]
SFG & 3--4 & 33 & $9.95^{+0.58}_{-0.45}$ & $1.21^{+0.51}_{-0.57}$ & $157^{+56}_{-42}$ & $76^{+27}_{-18}$ & $312^{+85}_{-61}$ & $7.00^{+0.12}_{-0.10}$ & $0.50^{+0.18}_{-0.15}$ & $41.0^{+0.4}_{-0.3}$ & $-0.71^{+0.74}_{-0.65}$ \\[4pt]
SFG & 4--5 & 44 & $9.40^{+0.58}_{-0.25}$ & $0.85^{+0.51}_{-0.51}$ & \nodata & \nodata & \nodata & \nodata & \nodata & \nodata & \nodata \\[4pt]
INT & 2--3 & 50 & $10.24^{+0.85}_{-0.46}$ & $0.47^{+0.71}_{-0.52}$ & $79^{+22}_{-21}$ & $75^{+24}_{-19}$ & $230^{+49}_{-41}$ & $7.05^{+0.15}_{-0.14}$ & $0.42^{+0.17}_{-0.16}$ & $40.7^{+0.3}_{-0.3}$ & $-0.09^{+0.75}_{-0.90}$ \\[4pt]
INT & 3--4 & 10 & $10.18^{+0.44}_{-0.43}$ & $0.49^{+0.52}_{-0.47}$ & \nodata & \nodata & \nodata & \nodata & \nodata & \nodata & \nodata \\[4pt]
INT & 4--5 & 14 & $9.97^{+0.42}_{-0.28}$ & $0.60^{+0.55}_{-0.24}$ & \nodata & \nodata & \nodata & \nodata & \nodata & \nodata & \nodata \\[4pt]
QG & 2--3 & 22 & $10.75^{+0.49}_{-0.48}$ & $-1.69^{+1.73}_{-3.90}$ & $105^{+34}_{-24}$ & $232^{+40}_{-34}$ & $574^{+77}_{-67}$ & $7.70^{+0.16}_{-0.14}$ & $1.47^{+0.15}_{-0.13}$ & $42.5^{+0.2}_{-0.2}$ & $3.45^{+5.99}_{-2.04}$ \\[4pt]
QG & 3--4 & 17 & $11.05^{+0.32}_{-0.37}$ & $-0.42^{+1.16}_{-2.92}$ & $126^{+29}_{-27}$ & $189^{+31}_{-29}$ & $505^{+54}_{-49}$ & $7.64^{+0.12}_{-0.09}$ & $1.36^{+0.10}_{-0.09}$ & $42.3^{+0.1}_{-0.1}$ & $1.68^{+1.64}_{-1.18}$ \\[4pt]
QG & 4--5 & 10 & $10.62^{+0.36}_{-0.45}$ & $-0.52^{+0.76}_{-3.14}$ & $128^{+39}_{-32}$ & $89^{+32}_{-24}$ & $313^{+64}_{-53}$ & $7.43^{+0.13}_{-0.11}$ & $0.94^{+0.16}_{-0.15}$ & $41.4^{+0.3}_{-0.3}$ & $1.45^{+4.21}_{-0.74}$ \\[4pt]
\enddata
\tablecomments{Columns are as follows: galaxy type, redshift bin, number of galaxies used in the stack ($N$), stellar mass, star formation rate, Na\,\textsc{i}\,D absorption velocity centroid, Na\,\textsc{i}\,D velocity dispersion, outflow velocity, neutral gas mass, mass outflow rate, kinetic energy outflow rate, and mass loading factor ($\eta \equiv \dot{M}_{\rm out}/{\rm SFR}$). This table shows the robustness test after removing galaxies with $M_\star < 10^9\,M_\odot$. The outflow velocity is defined as $v_{\rm out}=|v_{\rm Na\,\textsc{i}\,D}|+2\sigma_{\rm Na\,\textsc{i}\,D}$. Quoted uncertainties indicate the 16th, 50th, and 84th percentiles. Entries marked as \nodata\ correspond to Na\,\textsc{i}\,D non-detections or fits for which outflow properties could not be robustly constrained.}
\label{table:mass_test}
\end{deluxetable}

\begin{deluxetable}{lllccccccccc}
\tabletypesize{\scriptsize}
\tablewidth{0pt}
\tablecaption{Derived Physical and Outflow Properties for the High-mass Half Sample}
\tablehead{
\colhead{Type} &
\colhead{$z$} &
\colhead{$N$} &
\colhead{$\log M_\star$} &
\colhead{$\log {\rm SFR}$} &
\colhead{$|v_{\rm Na\,\textsc{i}\,D}|$} &
\colhead{$\sigma_{\rm Na\,\textsc{i}\,D}$} &
\colhead{$v_{\rm out}$} &
\colhead{$\log M_{\rm neutral}$} &
\colhead{$\log \dot{M}$} &
\colhead{$\log \dot{E}_{\rm out}$} &
\colhead{$\log \eta$} \\
\colhead{} &
\colhead{} &
\colhead{} &
\colhead{[$M_\odot$]} &
\colhead{[$M_\odot\,{\rm yr}^{-1}$]} &
\colhead{[km s$^{-1}$]} &
\colhead{[km s$^{-1}$]} &
\colhead{[km s$^{-1}$]} &
\colhead{[$M_\odot$]} &
\colhead{[$M_\odot\,{\rm yr}^{-1}$]} &
\colhead{[erg s$^{-1}$]} &
\colhead{}
}
\startdata
SFG & 2--3 & 25 & $10.36^{+0.66}_{-0.27}$ & $1.25^{+0.64}_{-0.30}$ & $102^{+41}_{-25}$ & $84^{+31}_{-23}$ & $275^{+78}_{-55}$ & $7.06^{+0.13}_{-0.14}$ & $0.51^{+0.18}_{-0.17}$ & $40.9^{+0.4}_{-0.3}$ & $-0.74^{+0.50}_{-0.79}$ \\[4pt]
SFG & 3--4 & 15 & $10.48^{+0.28}_{-0.42}$ & $1.47^{+0.55}_{-0.19}$ & $88^{+26}_{-22}$ & $89^{+29}_{-24}$ & $270^{+61}_{-54}$ & $7.17^{+0.14}_{-0.10}$ & $0.61^{+0.18}_{-0.16}$ & $41.0^{+0.3}_{-0.3}$ & $-0.83^{+0.36}_{-0.63}$ \\[4pt]
SFG & 4--5 & 6 & $10.26^{+0.20}_{-0.19}$ & $1.60^{+0.93}_{-0.49}$ & \nodata & \nodata & \nodata & \nodata & \nodata & \nodata & \nodata \\[4pt]
INT & 2--3 & 37 & $10.47^{+0.74}_{-0.30}$ & $0.85^{+0.41}_{-0.61}$ & $77.0^{+19}_{-19}$ & $77^{+24}_{-19}$ & $232^{+48}_{-39}$ & $7.32^{+0.13}_{-0.11}$ & $0.70^{+0.16}_{-0.15}$ & $40.9^{+0.3}_{-0.3}$ & $-0.15^{+0.91}_{-0.57}$ \\[4pt]
INT & 3--4 & 8 & $10.37^{+0.32}_{-0.32}$ & $0.84^{+0.22}_{-0.48}$ & \nodata & \nodata & \nodata & \nodata & \nodata & \nodata & \nodata \\[4pt]
INT & 4--5 & 8 & $10.33^{+0.43}_{-0.27}$ & $0.77^{+0.70}_{-0.42}$ & \nodata & \nodata & \nodata & \nodata & \nodata & \nodata & \nodata \\[4pt]
QG & 2--3 & 20 & $10.79^{+0.47}_{-0.40}$ & $-2.52^{+2.61}_{-3.52}$ & $118^{+36}_{-28}$ & $239^{+38}_{-33}$ & $599^{+75}_{-64}$ & $7.72^{+0.17}_{-0.14}$ & $1.52^{+0.14}_{-0.13}$ & $42.6^{+0.1}_{-0.2}$ & $3.33^{+4.43}_{-2.04}$ \\[4pt]
QG & 3--4 & 17 & $11.07^{+0.30}_{-0.36}$ & $-0.36^{+1.09}_{-3.12}$ & $126^{+30}_{-26}$ & $188^{+30}_{-30}$ & $501^{+54}_{-50}$ & $7.64^{+0.11}_{-0.10}$ & $1.35^{+0.10}_{-0.09}$ & $42.3^{+0.1}_{-0.1}$ & $1.64^{+3.89}_{-1.14}$ \\[4pt]
QG & 4--5 & 10 & $10.63^{+0.38}_{-0.44}$ & $-0.28^{+0.86}_{-3.70}$ & $130^{+42}_{-37}$ & $92^{+34}_{-26}$ & $319^{+69}_{-56}$ & $7.42^{+0.13}_{-0.12}$ & $0.93^{+0.18}_{-0.16}$ & $41.4^{+0.3}_{-0.3}$ & $1.62^{+6.42}_{-1.38}$ \\[4pt]
\enddata
\tablecomments{Columns are as follows: galaxy type, redshift bin, number of stacked galaxies ($N$), stellar mass, star formation rate, Na\,\textsc{i}\,D absorption velocity centroid, Na\,\textsc{i}\,D velocity dispersion, outflow velocity, neutral gas mass, mass outflow rate, kinetic energy outflow rate, and mass loading factor ($\eta \equiv \dot{M}_{\rm out}/{\rm SFR}$). This table uses the high-mass half robustness-test sample selected by $\log M_\star \geq 9.9926$. Outflow properties are derived from the Na\,\textsc{i}\,D absorption-line stacking and fitting. The outflow velocity is defined as $v_{\rm out}=|v_{\rm NaD}|+2\sigma_{\rm NaD}$. Quoted uncertainties indicate the 16th, 50th, and 84th percentiles. Entries marked as \nodata\ correspond to Na\,\textsc{i}\,D non-detections or unconstrained outflow properties.}
\label{table:mass_high_half}
\end{deluxetable}

\begin{deluxetable}{lllccccccccc}
\tabletypesize{\scriptsize}
\tablewidth{0pt}
\tablecaption{Derived Physical and Outflow Properties for the Low-mass Half Sample}
\tablehead{
\colhead{Type} &
\colhead{$z$} &
\colhead{$N$} &
\colhead{$\log M_\star$} &
\colhead{$\log {\rm SFR}$} &
\colhead{$|v_{\rm Na\,\textsc{i}\,D}|$} &
\colhead{$\sigma_{\rm Na\,\textsc{i}\,D}$} &
\colhead{$v_{\rm out}$} &
\colhead{$\log M_{\rm neutral}$} &
\colhead{$\log \dot{M}$} &
\colhead{$\log \dot{E}_{\rm out}$} &
\colhead{$\log \eta$} \\
\colhead{} &
\colhead{} &
\colhead{} &
\colhead{[$M_\odot$]} &
\colhead{[$M_\odot\,{\rm yr}^{-1}$]} &
\colhead{[km s$^{-1}$]} &
\colhead{[km s$^{-1}$]} &
\colhead{[km s$^{-1}$]} &
\colhead{[$M_\odot$]} &
\colhead{[$M_\odot\,{\rm yr}^{-1}$]} &
\colhead{[erg s$^{-1}$]} &
\colhead{}
}
\startdata
SFG & 2--3 & 30 & $9.67^{+0.19}_{-0.35}$ & $0.79^{+0.27}_{-0.34}$ & \nodata & \nodata & \nodata & \nodata & \nodata & \nodata & \nodata \\[4pt]
SFG & 3--4 & 19 & $9.55^{+0.30}_{-0.21}$ & $1.05^{+0.17}_{-0.49}$ & \nodata & \nodata & \nodata & \nodata & \nodata & \nodata & \nodata \\[4pt]
SFG & 4--5 & 56 & $9.18^{+0.41}_{-0.46}$ & $0.77^{+0.34}_{-0.46}$ & \nodata & \nodata & \nodata & \nodata & \nodata & \nodata & \nodata \\[4pt]
INT & 2--3 & 13 & $9.73^{+0.16}_{-0.29}$ & $-0.16^{+0.30}_{-0.12}$ & \nodata & \nodata & \nodata & \nodata & \nodata & \nodata & \nodata \\[4pt]
INT & 3--4 & 2 & $9.25^{+0.44}_{-0.02}$ & $0.13^{+0.17}_{-0.31}$ & \nodata & \nodata & \nodata & \nodata & \nodata & \nodata & \nodata \\[4pt]
INT & 4--5 & 6 & $9.88^{+0.06}_{-0.51}$ & $0.60^{+0.03}_{-0.20}$ & \nodata & \nodata & \nodata & \nodata & \nodata & \nodata & \nodata \\[4pt]
QG & 2--3 & 2 & $9.70^{+0.03}_{-0.03}$ & $-0.72^{+0.20}_{-0.20}$ & \nodata & \nodata & \nodata & \nodata & \nodata & \nodata & \nodata \\[4pt]
QG & 3--4 & 0 & \nodata & \nodata & \nodata & \nodata & \nodata & \nodata & \nodata & \nodata & \nodata \\[4pt]
QG & 4--5 & 0 & \nodata & \nodata & \nodata & \nodata & \nodata & \nodata & \nodata & \nodata & \nodata \\[4pt]
\enddata
\tablecomments{Columns are as follows: galaxy type, redshift bin, number of galaxies included in the Na\,\textsc{i}\,D stack ($N$), stellar mass, star formation rate, Na\,\textsc{i}\,D absorption velocity centroid, Na\,\textsc{i}\,D velocity dispersion, outflow velocity, neutral gas mass, mass outflow rate, kinetic energy outflow rate, and mass loading factor ($\eta \equiv \dot{M}_{\rm out}/{\rm SFR}$). This table uses the low-mass half robustness-test sample selected by $\log M_\star < 9.9926$. Outflow properties are reported only when the Na\,\textsc{i}\,D stack and subsequent fitting are robustly constrained. No low-mass QGs are present at $3 \leq z < 4$ or $4 \leq z < 5$ in the parent low-mass sample; no Na\,\textsc{i}\,D stack was performed for these bins, and all corresponding entries are therefore reported as \nodata. The low-mass QG bin at $2 \leq z < 3$ and the low-mass INT bin at $3 \leq z < 4$ have very small sample sizes, so their outflow properties are not interpreted. Quoted uncertainties indicate the 16th, 50th, and 84th percentiles. Entries marked as \nodata\ correspond to bins without a performed stack, Na\,\textsc{i}\,D non-detections, insufficient sample size, or unconstrained outflow properties.}
\label{table:mass_low_half}
\end{deluxetable}

\begin{deluxetable}{lll}
\tablecaption{Uniform priors adopted for the absorption line fitting.\label{tab:naid_priors}}
\tablehead{
\colhead{Parameter} & \colhead{Prior range} & \colhead{Description}
}
\startdata
$C_f$ & $[0, 1]$ & Covering fraction \\
$\log\tau_{0,r}$ & $[-2.0, 0.8]$ & Optical depth of the red Na\,\textsc{i}\,D component \\
$v_{\rm NaD}$ & $[-600, 200]~{\rm km~s^{-1}}$ & Velocity offset of the Na\,\textsc{i}\,D absorption \\
$\sigma_{\rm NaD}$ & $[35, 500]~{\rm km~s^{-1}}$ & Na\,\textsc{i}\,D velocity dispersion \\
continuum & $[-0.05, 0.05]$ & Continuum baseline offset \\
$A_{\rm He\,\textsc{i},emi}$ & $[0, 0.6]$ & Amplitude of the He\,\textsc{i} emission component \\
$\Delta v_{\rm He\,\textsc{i},emi}$ & $[-150, 150]~{\rm km~s^{-1}}$ & Velocity offset of the He\,\textsc{i} emission component \\
$A_{\rm He\,\textsc{i},abs}$ & $[-0.3, 0]$ & Amplitude of the He\,\textsc{i} absorption component \\
$\Delta v_{\rm He\,\textsc{i},abs}$ & $[-600, -100]~{\rm km~s^{-1}}$ & Velocity offset of the He\,\textsc{i} absorption component \\
\enddata
\tablecomments{Columns are as follows: model parameter, adopted prior range, and parameter description. All priors are uniform within the listed bounds. The Na\,\textsc{i}\,D absorption profile is modeled using a partial-covering doublet model, with $C_f$, $\tau_{0,r}$, $v_{\rm NaD}$, $\sigma_{\rm NaD}$, and continuum describing the covering fraction, red-component optical depth, velocity offset relative to systemic, velocity dispersion, and the residual baseline after subtraction, respectively. The He\,\textsc{i} emission and absorption components are included to account for possible P-Cygni-like contamination near the Na\,\textsc{i}\,D doublet. The adopted ranges are broad but exclude unphysical values and unrealistically broad solutions.
}
\end{deluxetable}

\end{document}